\documentclass[runningheads]{cl2emult}

\usepackage{harvard}
\usepackage{amssymb}
\usepackage{latexsym}

\usepackage{xy}
\xyoption{all}
\CompileMatrices 
\UseComputerModernTips

\def\Onestep{\ar@{>}}
\def\Manyred{\ar@{>>}}
\def\Emptystep{\ar@{.}}
\def\Symstep{\ar@{|-|}}
\def\Line{\ar@{-}}
\def\Noline{\ar@{}}
\def\Dashed{\ar@{--}}
\def\Manydots{\ar@{.>>}}
\def\Onedots{\ar@{.>}}
\def\Linedots{\ar@{.}}
\def\Curldash{\ar@{~~}}
\def\Linecurl{\ar@{~}}
\def\Doubline{\ar@{=}}

\newcommand{\ra}{\mbox{$\:\rightarrow\:$}}

\newcommand{\lra}{\mbox{$\:\leftrightarrow\:$}}

\newcommand{\A}{\mbox{$\ \wedge\ $}}

\newcommand{\Or}{\mbox{$\ \vee\ $}}

\newcommand{\fa}{\mbox{$\forall$}}
\newcommand{\te}{\mbox{$\exists$}}

\newcommand{\LL}{\mbox{$\ldots$}}

\newcommand{\B}[1]{\mbox{$[\![{#1}]\!]$}}       
\newcommand{\C}[1]{\mbox{$\{{#1}\}$}}           

\newcommand{\NI}{\noindent}
\newcommand{\HB}{\hfill{$\boxtimes$}}
\newcommand{\VV}{\vspace{5 mm}}

\newcounter{oldmycaption}

\def\smallromani{\renewcommand{\theenumi}{\roman{enumi}}
\renewcommand{\labelenumi}{(\theenumi)}}

\newcommand{\Proof}{\NI
                    {\bf Proof.}\ }

\newcounter{symbol}
\newcommand{\indexsyma}[1]%
{\stepcounter{symbol}\index{zzz1 \thesymbol @\protect#1}}
\newcommand{\indexsymb}[1]%
{\stepcounter{symbol}\index{zzz2 \thesymbol @\protect#1}}
\newcommand{\indexsymc}[1]%
{\stepcounter{symbol}\index{zzz3 \thesymbol @\protect#1}}
\newcommand{\indexsymd}[1]%
{\stepcounter{symbol}\index{zzz4 \thesymbol @\protect#1}}
\newcommand{\indexsyme}[1]%
{\stepcounter{symbol}\index{zzz5 \thesymbol @\protect#1}}

\newcommand{\almazero}{\mbox{\sf Alma-0}}
\def\err{\mathit{error}}
\def\fal{\mathit{fail}}

\def\empc{\Box}
\def\empval{\varepsilon}

\begin{document}

\title{Formulas as Programs}
\def\tensfi{\em}
\institute{CWI, P.O. Box 94079, 1090 GB Amsterdam, The Netherlands}
\author{Krzysztof R. Apt
\and Marc Bezem
}

\maketitle

\begin{abstract}
  We provide here a computational interpretation of first-order logic
  based on a constructive interpretation of satisfiability w.r.t. a
  fixed but arbitrary interpretation. In this approach the
  \emph{formulas} themselves are \emph{programs}. This contrasts with
  the so-called \emph{formulas as types} approach in which the proofs
  of the  formulas are typed terms that can be taken as programs.
  This view of computing is inspired by logic programming 
  and constraint logic programming but differs
  from them in a number of crucial aspects.

  Formulas as programs is argued to yield a realistic approach to
  programming that has been realized in the implemented programming
  language \almazero{} \citeasnoun{ABPS98a} that combines the advantages of
  imperative and logic programming.  The work here
  reported can also be used to reason about the correctness of 
  non-recursive \almazero{}
  programs that do not include destructive assignment.
\end{abstract}


\section{Introduction}

\subsection{Logic Programming and Program Verification}

The logic programming paradigm in its original form
(see \citeasnoun{Kow74}) is based on a computational 
interpretation of a subset of first-order logic that 
consists of Horn clauses.
The proof theory and semantics for this subset has been
well understood for some time already (see, e.g. \citeasnoun{Llo87}).

However, the practice has quickly shown that this subset is too
limited for the programming purposes, so 
it was extended in a number of ways, notably by allowing
negation. This led to a long and still inconclusive quest for
extending the appropriate soundness and completeness results
to logic programs that allow negation
(see, e.g. \citeasnoun{AB94}).
To complicate the matters further, Prolog extends logic programming
with negation by several features that are very operational
in nature.

Constraint logic programming (see, e.g. \citeasnoun{jaffar-constraint-87})
overcomes some of Prolog's deficiencies, notably its clumsy handling
of arithmetic, by extending the computing process 
from the (implicit) domain of terms to arbitrary structures.

Logic programming and constraint logic programming are two 
instances of declarative programming. According to
declarative programming a program has
a dual reading as a formula in a logic with a simple semantics.

One of the important advantages of
declarative programming is that, thanks to the semantic
interpretation, 
programs are easier to understand, modify and verify.
In fact, the dual reading of a declarative program as a formula allows
us to reason about its correctness by restricting our attention to a
logical analysis of the corresponding formula.  For each logical
formalism such an analysis essentially boils down to the question
whether the formula corresponding to the program is in an appropriate
sense equivalent to the specification.\footnote{
This can be made precise in the following way. Let $\vec{x}$ be the
free variables of the specification $\phi_s$, and $\vec{y}$ some auxiliary
variables used in the program $\phi_p$. Now correctness of the program
with respect to the specification can be expressed by the sentence
$\fa\vec{x}~((\te\vec{y}~\phi_p(\vec{x},\vec{y})) \ra \phi_s(\vec{x}))$,
to be valid under the fixed interpretation. This sentence ensures
that all solutions found by the program indeed satisfy the specification.
Note that, under this definition, a program corresponding to a false
formula is vacuously ``correct'', because there are no solutions found.
Therefore the stronger notion of correctness and completeness obtained
by requiring also the converse implication above, and loosely phrased as
``equivalence in an appropriate sense'', is the more adequate one.}

However, in our opinion, we do not have at our disposal
{\em simple\/} and {\em intuitive\/} methods that could be 
used to verify in a rigorous way realistic ``pure'' Prolog programs
(i.e. those that are also logic programs) or
constraint logic programs.

We believe that one of the reasons for this state of affairs is
recursion, on which both logic programming and constraint logic 
programming rely. 
In fact, recursion is often less natural than iteration,
which is a more basic concept.
Further, 
recursion in combination with negation can naturally lead to programs 
that are not easily amenable to a formal analysis.
Finally, recursion always introduces a possibility of divergence
which explains why the study of termination is such an important
topic in the case of logic programming (see, e.g., \citeasnoun{DD94}).

\subsection{First-order Logic as a Computing Mechanism}

Obviously, without recursion logic programming and constraint 
logic programming are hopelessly inexpressive. However, as we 
show in this paper, it is still possible to construct a 
simple and realistic
approach to declarative programming that draws on the ideas of these
two formalisms and in which recursion is absent.
This is done by providing a constructive interpretation of 
satisfiability of first-order formulas w.r.t. to a fixed but 
arbitrary interpretation. Iteration is realized by means
of bounded quantification that is guaranteed to terminate.

More precisely, assuming a first-order language $L$, we introduce an
effective, though incomplete, computation mechanism that approximates
the satisfiability test in the following sense.  Given an
interpretation $I$ for $L$ and a formula $\phi(\bar{x})$ of $L$, 
assuming no abnormal termination in an error arises, this mechanism
computes a witness $\bar{a}$ (that is, a vector of elements of the
domain of $I$ such that $\phi(\bar{a})$ holds in $I$) if
$\phi(\bar{x})$ is satisfiable in $I$, and otherwise it reports a
failure.

The possibility of abnormal termination in an error is unavoidable
because effectiveness cannot be reconciled with the fact that for many
first-order languages and interpretations, for example the language of
Peano arithmetic and its standard interpretation, the set of true
closed formulas is highly undecidable.  As we wish to use this
computation mechanism for executing formulas as programs, we spend
here considerable effort at investigating the ways of limiting the
occurrence of errors.

From the technical point of view our approach, 
called {\em formulas as programs\/}, is obtained by
isolating a number of concepts and ideas present (often implicitly)
in the logic programming and constraint logic programming
framework, and reusing them in a simple and self-contained way.
In fact, the proposed computation mechanism and a rigorous account
of its formal properties rely only on the
basics of first-order logic. This contrasts
with the expositions of logic programming and constraint
logic programming which require
introduction of several concepts and auxiliary results
(see for the latter e.g. \citeasnoun{JMMS98}).

\subsection{Computing Mechanism}
\label{ssec:informal}

Let us explain now the proposed computation mechanism by
means of an example.
Consider the formula
\begin{equation}
(x = 2 \Or x = 3)\A (y = x + 1 \Or 2 = y) \A (2*x = 3*y)
  \label{eq:no-prolog}
\end{equation}
interpreted over the standard structure of natural numbers.
Is it satisfiable? The answer is ``yes'': indeed, it suffices
to assign 3 to $x$ and 2 to $y$.

In fact, we can compute this valuation systematically by initially
assigning 2 to $x$ and first trying the assignment of the value of
$x+1$, so 3, to $y$.  As
for this choice of value for $y$ the equality $2*x = 3*y$ does not
hold, we are led to the second possibility, assignment of 2 to $y$.
With this choice $2*x = 3*y$ does not hold either. So we need to
assign 3 to $x$ and, eventually, 2 to $y$.

The above informal argument can be extended to a systematic
procedure that attempts to find a satisfying valuation
for a large class of formulas.

\subsection{Plan and Rationale of the Paper}
\label{ssec:rationale}
This paper is organized as follows.

In Section~\ref{sec:compu} we provide a
formal account of the proposed computation mechanism.
In Section~ \ref{sec:souncom}
we show that this approach is both
correct (sound) and, in the absence of errors, complete.
In the Appendix, Subsection~\ref{ssec:libneg}, \ref{ssec:libimp},
we investigate ways of limiting the occurrence
of errors for the case of negation and implication.

For programming purposes first-order logic has limited expressiveness,
so we extend it in Section~\ref{sec:extensions} by a number of
features that are useful for programming. This involves sorts (i.e.,
types), use of arrays and bounded quantifiers. The resulting fragment
is surprisingly expressive and the underlying computation mechanism
allows us to interpret many formulas as highly non-trivial programs.

As already mentioned above, formulas as programs approach to 
computing here discussed is
inspired by logic programming and constraint logic programming
but differs from them in a number of ways.

For example, formula (\ref{eq:no-prolog}) cannot be interpreted as a
logic programming query or run as a Prolog query.  The reason is that
the equality symbol in logic programming and Prolog stands for ``is
unifiable with'' and the term $2*x$ does not unify with $3*y$.  In
case of Prolog a possible remedy is to replace in (\ref{eq:no-prolog})
specific occurrences of the equality symbol by Prolog's arithmetic
equality ``\verb|=:=|'' or by the Prolog evaluator operator {\tt is}.  The
correct Prolog query that corresponds to formula (\ref{eq:no-prolog})
is then

{\tt (X = 2 ; X = 3), (Y is X+1 ; 2 = Y), 2*X =:= 3*Y.}

\NI
(Recall that ``\verb|;|'' stands in Prolog for disjunction and 
``\verb|,|'' for
conjunction.)  This is clearly much less readable than
(\ref{eq:no-prolog}) as three different kinds of equality-like
relations are used here.

A more detailed comparison with (constraint) logic programming and
Prolog requires knowledge of the details of our approach and is
postponed to Section~\ref{sec:related}.  In principle, the formulas
as programs approach is a variant of constraint logic programming
in which both recursion and constraint handling procedures are absent,
but the full first-order syntax is used.
We also compare in Section~\ref{sec:related}  our
formulas as programs approach with the formulas as types approach,
also called the Curry-Howard-De Bruijn interpretation.

The formulas as programs 
approach to programming has been realized in the programming
language \almazero{} \citeasnoun{ABPS98a} that extends imperative
programming by features that support declarative programming.
This shows that this approach, in contrast to logic programming
and constraint logic programming, can easily be 
combined with imperative
programming. So the introduced restrictions, such
as lack of a constraint store, can be beneficial in practice.
In Section~\ref{sec:alma0} we summarize the main features of
\almazero{}.

The work reported here can be used to provide logical underpinnings
for a fragment of \almazero{} that does not include destructive
assignment or recursive procedures,
and to reason about programs written in this fragment.
We substantiate the latter claim by presenting in
Section~\ref{sec:squares} the correctness proof of a purely declarative
\almazero{} solution to the well-known non-trivial combinatorial problem
of partitioning a rectangle into a given set of squares.

In conclusion, we provided here 
a realistic framework for declarative programming
based on first-order logic and the traditional 
Tarskian semantics, which can be combined in a straightforward
way with imperative programming.

\section{Computation Mechanism}
\label{sec:compu}
Consider an arbitrary first-order language with equality
and an interpretation for it.
We assume in particular a domain of discourse,
and a fixed signature with a
corresponding interpretation of its elements in the domain.
\begin{definition}[valuation, assignment]\label{def:valuation}
A \emph{valuation} is a finite mapping from variables to domain elements.
Valuations will be denoted as single-valued sets of pairs
$x/d$, where $x$ is a variable and $d$ a domain element.
We use $\alpha,\alpha',\beta,\ldots$ for arbitrary valuations and
call $\alpha'$ an \emph{extension} of $\alpha$ when $\alpha\subseteq\alpha'$,
that is, every assignment to a variable by $\alpha$ also occurs in $\alpha'$.
Further, $\empval$ denotes the empty valuation.

Let $\alpha$ be a valuation.
A term $t$ is $\alpha$-\emph{closed} if all variables of $t$ get a value
in $\alpha$. In that case $t^\alpha$ denotes the \emph{evaluation}
of $t$ under $\alpha$ in the domain.
More generally, for any expression $E$ the result of the replacement
of each $\alpha$-closed term $t$ by $t^\alpha$ is denoted by $E^\alpha$.

An $\alpha$-assignment is an equation $s = t$ one side of which,
say  $s$, is a variable that is not $\alpha$-closed and the other
side, $t$, is an $\alpha$-closed term.\HB
\end{definition}

In our setting, the only way to assign values to variables
will be by evaluating an $\alpha$-assignment as above.
Given such an $\alpha$-assignment, say $x = t$, we evaluate it
by assigning to $x$ the value $t^\alpha$.

\begin{definition}[formulas]\label{def:formulas}
In order to accommodate the definition of the operational semantics,
the set of fomulas has an inductive definition which may look a bit
peculiar. First, universal quantification is absent since we
have no operational interpretation for it.
Second, every formula is taken to be
a conjunction, with every conjunct (if any) either
an atomic formula (in short: an
{\em atom\/}), or a disjunction, conjunction or implication of formulas, 
a negation of a formula or an existentially quantified formula. 
The latter two unary constructors are assumed to bind stronger
then the previous binary ones.
The atoms include equations of the form $s=t$, with $s$ and $t$ terms.

For maximal clarity we
give here an inductive definition of the set of formulas.
In the operational semantics all conjunctions are taken to
be right associative.
\begin{enumerate}
\item The empty conjunction $\empc$ is a formula.
\item If $\psi$ is a formula and $A$ is an atom, then $A\wedge\psi$
is a formula.
\item If $\psi,\phi_1,\phi_2$ are formulas, 
then $(\phi_1\vee\phi_2)\wedge\psi$ is a formula.
\item If $\psi,\phi_1,\phi_2$ are formulas, 
then $(\phi_1\wedge\phi_2)\wedge\psi$ is a formula.
\item If $\psi,\phi_1,\phi_2$ are formulas, 
then $(\phi_1 \ra \phi_2)\wedge\psi$ is a formula.
\item If $\phi,\psi$ are formulas, 
then $\neg\phi \wedge\psi$ is a formula.
\item If $\phi,\psi$ are formulas, 
then $\exists x\;\phi \wedge\psi$ is a formula.
\HB
\end{enumerate}
\end{definition}

\begin{definition}[operational semantics]\label{def:opersem}
The operational semantics of a formula will be defined in terms
of a tree $\B{\phi}_{\alpha}$ depending on the formula $\phi$
and the (initial) valuation $\alpha$. 
The root of $\B{\phi}_{\alpha}$ is labelled with the pair $\phi,\alpha$.
All internal nodes of the tree $\B{\phi}_{\alpha}$
are labelled with pairs consisting of a formula and a valuation.
The leaves of the tree $\B{\phi}_{\alpha}$
are labelled with either 
\begin{itemize}
\item $\err$ (representing the occurrence
of an error in this branch of the computation), or 
\item $\fal$ (representing
logical failure of the computation), or 
\item a valuation (representing
logical success of the computation and yielding values for the
free variables of the formula that make the formula true).~$\boxtimes$
\end{itemize}
\end{definition}

It will be shown that valuations labelling success leaves are
always extensions of the initial valuation.
For a fixed formula, the operational semantics can be viewed as a function
relating the initial valuation to the valuations labelling success leaves.

We can now define the computation tree $\B{\phi}_{\alpha}$.
The reader may consult first
Fig.~\ref{fig:ctree} to see such a tree for formula
(\ref{eq:no-prolog}) and the empty valuation $\empval$.

\begin{figure}[htbp]
\begin{center}
\input{aptbezem1a.pstex_t}
\caption{The computation tree for formula (1) and valuation $\empval$. 
\label{fig:ctree}}
\end{center}

\end{figure}

\begin{definition}[computation tree]\label{def:tree}
The (computation) tree $\B{\phi}_{\alpha}$ is defined by 
lexicographic induction on the pairs consisting of
the \emph{size} of the formula $\phi$, and of the
\emph{size} of the formula $\phi_1$ for which
$\phi$ is of the form $\phi_1 \A \psi$,
following the
structure given by Definition~\ref{def:formulas}.
\begin{enumerate}

\item\label{os:empty}
For the empty conjunction we define $\B{\empc}_\alpha$ 
to be the tree with the root that has
a success leaf $\alpha$ as its son:

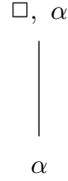
\begin{center}
\begin{picture}(0,0)%
\includegraphics{aptbezem2.pstex}%
\end{picture}%
\setlength{\unitlength}{3947sp}%
\begingroup\makeatletter\ifx\SetFigFont\undefined
\def\x#1#2#3#4#5#6#7\relax{\def\x{#1#2#3#4#5#6}}%
\expandafter\x\fmtname xxxxxx\relax \def\y{splain}%
\ifx\x\y   
\gdef\SetFigFont#1#2#3{%
  \ifnum #1<17\tiny\else \ifnum #1<20\small\else
  \ifnum #1<24\normalsize\else \ifnum #1<29\large\else
  \ifnum #1<34\Large\else \ifnum #1<41\LARGE\else
     \huge\fi\fi\fi\fi\fi\fi
  \csname #3\endcsname}%
\else
\gdef\SetFigFont#1#2#3{\begingroup
  \count@#1\relax \ifnum 25<\count@\count@25\fi
  \def\x{\endgroup\@setsize\SetFigFont{#2pt}}%
  \expandafter\x
    \csname \romannumeral\the\count@ pt\expandafter\endcsname
    \csname @\romannumeral\the\count@ pt\endcsname
  \csname #3\endcsname}%
\fi
\fi\endgroup
\begin{picture}(1000,1140)(2501,-919)
\put(3001, 89){\makebox(0,0)[b]{\smash{\SetFigFont{10}{13.2}{rm}$\Box, \ \alpha$}}}
\put(3001,-886){\makebox(0,0)[b]{\smash{\SetFigFont{10}{13.2}{rm}$\alpha$}}}
\end{picture}

\end{center}

\item\label{os:atom}
If $\psi$ is a formula and $A$ is an atom, 
then we distinguish four cases depending on the form of $A$.
In all four cases $\B{A\wedge\psi}_\alpha$ is a tree
with a root of degree one.
\begin{itemize}
\item Atom $A$ is $\alpha$-closed and true. 
Then the root of $\B{A\wedge\psi}_\alpha$ has $\B{\psi}_\alpha$ as its subtree:

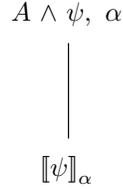
\begin{center}
\begin{picture}(0,0)%
\includegraphics{aptbezem3.pstex}%
\end{picture}%
\setlength{\unitlength}{3947sp}%
\begingroup\makeatletter\ifx\SetFigFont\undefined
\def\x#1#2#3#4#5#6#7\relax{\def\x{#1#2#3#4#5#6}}%
\expandafter\x\fmtname xxxxxx\relax \def\y{splain}%
\ifx\x\y   
\gdef\SetFigFont#1#2#3{%
  \ifnum #1<17\tiny\else \ifnum #1<20\small\else
  \ifnum #1<24\normalsize\else \ifnum #1<29\large\else
  \ifnum #1<34\Large\else \ifnum #1<41\LARGE\else
     \huge\fi\fi\fi\fi\fi\fi
  \csname #3\endcsname}%
\else
\gdef\SetFigFont#1#2#3{\begingroup
  \count@#1\relax \ifnum 25<\count@\count@25\fi
  \def\x{\endgroup\@setsize\SetFigFont{#2pt}}%
  \expandafter\x
    \csname \romannumeral\the\count@ pt\expandafter\endcsname
    \csname @\romannumeral\the\count@ pt\endcsname
  \csname #3\endcsname}%
\fi
\fi\endgroup
\begin{picture}(1277,1140)(2662,-919)
\put(3301, 89){\makebox(0,0)[b]{\smash{\SetFigFont{10}{13.2}{rm}$A \A \psi, \ \alpha$}}}
\put(3301,-886){\makebox(0,0)[b]{\smash{\SetFigFont{10}{13.2}{rm}$\B{\psi}_\alpha$}}}
\end{picture}

\end{center}

\item Atom $A$ is $\alpha$-closed and false.
Then the root of $\B{A\wedge\psi}_\alpha$ has the failure leaf $\fal$
as its son:

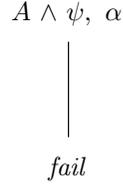
\begin{center}
\begin{picture}(0,0)%
\includegraphics{aptbezem4.pstex}%
\end{picture}%
\setlength{\unitlength}{3947sp}%
\begingroup\makeatletter\ifx\SetFigFont\undefined
\def\x#1#2#3#4#5#6#7\relax{\def\x{#1#2#3#4#5#6}}%
\expandafter\x\fmtname xxxxxx\relax \def\y{splain}%
\ifx\x\y   
\gdef\SetFigFont#1#2#3{%
  \ifnum #1<17\tiny\else \ifnum #1<20\small\else
  \ifnum #1<24\normalsize\else \ifnum #1<29\large\else
  \ifnum #1<34\Large\else \ifnum #1<41\LARGE\else
     \huge\fi\fi\fi\fi\fi\fi
  \csname #3\endcsname}%
\else
\gdef\SetFigFont#1#2#3{\begingroup
  \count@#1\relax \ifnum 25<\count@\count@25\fi
  \def\x{\endgroup\@setsize\SetFigFont{#2pt}}%
  \expandafter\x
    \csname \romannumeral\the\count@ pt\expandafter\endcsname
    \csname @\romannumeral\the\count@ pt\endcsname
  \csname #3\endcsname}%
\fi
\fi\endgroup
\begin{picture}(1277,1134)(2362,-913)
\put(3001, 89){\makebox(0,0)[b]{\smash{\SetFigFont{10}{13.2}{rm}$A \A \psi, \ \alpha$}}}
\put(3001,-886){\makebox(0,0)[b]{\smash{\SetFigFont{10}{13.2}{rm}{\em fail}}}}
\end{picture}

\end{center}

\item Atom $A$ is not $\alpha$-closed, but is not an $\alpha$-assignment.
Then the root of $\B{A\wedge\psi}_\alpha$ has the $\err$ leaf as its son:
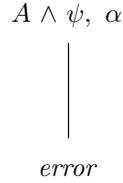
\begin{center}
\begin{picture}(0,0)%
\includegraphics{aptbezem5.pstex}%
\end{picture}%
\setlength{\unitlength}{3947sp}%
\begingroup\makeatletter\ifx\SetFigFont\undefined
\def\x#1#2#3#4#5#6#7\relax{\def\x{#1#2#3#4#5#6}}%
\expandafter\x\fmtname xxxxxx\relax \def\y{splain}%
\ifx\x\y   
\gdef\SetFigFont#1#2#3{%
  \ifnum #1<17\tiny\else \ifnum #1<20\small\else
  \ifnum #1<24\normalsize\else \ifnum #1<29\large\else
  \ifnum #1<34\Large\else \ifnum #1<41\LARGE\else
     \huge\fi\fi\fi\fi\fi\fi
  \csname #3\endcsname}%
\else
\gdef\SetFigFont#1#2#3{\begingroup
  \count@#1\relax \ifnum 25<\count@\count@25\fi
  \def\x{\endgroup\@setsize\SetFigFont{#2pt}}%
  \expandafter\x
    \csname \romannumeral\the\count@ pt\expandafter\endcsname
    \csname @\romannumeral\the\count@ pt\endcsname
  \csname #3\endcsname}%
\fi
\fi\endgroup
\begin{picture}(1277,1134)(2362,-913)
\put(3001,-886){\makebox(0,0)[b]{\smash{\SetFigFont{10}{13.2}{rm}{\em error}}}}
\put(3001, 89){\makebox(0,0)[b]{\smash{\SetFigFont{10}{13.2}{rm}$A \A \psi, \ \alpha$}}}
\end{picture}

\end{center}

\item Atom $A$ is an $\alpha$-assignment $s = t$. Then either
$s$ or $t$ is a variable which is not $\alpha$-closed,
say $s \equiv x$ with $x$ not $\alpha$-closed and $t$
$\alpha$-closed. Then the root of $\B{A\wedge\psi}_\alpha$ has
$\B{\psi}_{\alpha'}$ as its subtree, where $\alpha'$ extends
$\alpha$ with the pair $x/t^\alpha$:
\begin{center}
\begin{picture}(0,0)%
\includegraphics{aptbezem6.pstex}%
\end{picture}%
\setlength{\unitlength}{3947sp}%
\begingroup\makeatletter\ifx\SetFigFont\undefined
\def\x#1#2#3#4#5#6#7\relax{\def\x{#1#2#3#4#5#6}}%
\expandafter\x\fmtname xxxxxx\relax \def\y{splain}%
\ifx\x\y   
\gdef\SetFigFont#1#2#3{%
  \ifnum #1<17\tiny\else \ifnum #1<20\small\else
  \ifnum #1<24\normalsize\else \ifnum #1<29\large\else
  \ifnum #1<34\Large\else \ifnum #1<41\LARGE\else
     \huge\fi\fi\fi\fi\fi\fi
  \csname #3\endcsname}%
\else
\gdef\SetFigFont#1#2#3{\begingroup
  \count@#1\relax \ifnum 25<\count@\count@25\fi
  \def\x{\endgroup\@setsize\SetFigFont{#2pt}}%
  \expandafter\x
    \csname \romannumeral\the\count@ pt\expandafter\endcsname
    \csname @\romannumeral\the\count@ pt\endcsname
  \csname #3\endcsname}%
\fi
\fi\endgroup
\begin{picture}(1277,1140)(2662,-919)
\put(3301,-886){\makebox(0,0)[b]{\smash{\SetFigFont{10}{13.2}{rm}$\B{\psi}_{\alpha'}$}}}
\put(3301, 89){\makebox(0,0)[b]{\smash{\SetFigFont{10}{13.2}{rm}$A \A \psi, \ \alpha$}}}
\end{picture}

\end{center}

The symmetrical case is analogous.
\end{itemize}

\item\label{os:disj}
If $\psi,\phi_1,\phi_2$ are formulas,
then we put $\B{(\phi_1\vee\phi_2)\wedge\psi}_\alpha$ to be 
the tree with a root of degree two and with left and right subtrees
$\B{\phi_1\wedge\psi}_\alpha$ and
$\B{\phi_2\wedge\psi}_\alpha$, respectively:

\begin{center}
\begin{picture}(0,0)%
\includegraphics{aptbezem7.pstex}%
\end{picture}%
\setlength{\unitlength}{3947sp}%
\begingroup\makeatletter\ifx\SetFigFont\undefined
\def\x#1#2#3#4#5#6#7\relax{\def\x{#1#2#3#4#5#6}}%
\expandafter\x\fmtname xxxxxx\relax \def\y{splain}%
\ifx\x\y   
\gdef\SetFigFont#1#2#3{%
  \ifnum #1<17\tiny\else \ifnum #1<20\small\else
  \ifnum #1<24\normalsize\else \ifnum #1<29\large\else
  \ifnum #1<34\Large\else \ifnum #1<41\LARGE\else
     \huge\fi\fi\fi\fi\fi\fi
  \csname #3\endcsname}%
\else
\gdef\SetFigFont#1#2#3{\begingroup
  \count@#1\relax \ifnum 25<\count@\count@25\fi
  \def\x{\endgroup\@setsize\SetFigFont{#2pt}}%
  \expandafter\x
    \csname \romannumeral\the\count@ pt\expandafter\endcsname
    \csname @\romannumeral\the\count@ pt\endcsname
  \csname #3\endcsname}%
\fi
\fi\endgroup
\begin{picture}(3061,1215)(2370,-1294)
\put(3901,-211){\makebox(0,0)[b]{\smash{\SetFigFont{10}{13.2}{rm}$(\phi_1 \Or \phi_2) \A \psi, \ \alpha$}}}
\put(4501,-1261){\makebox(0,0)[b]{\smash{\SetFigFont{10}{13.2}{rm}$\B{\phi_1 \A \psi}_{\alpha}$}}}
\put(3301,-1261){\makebox(0,0)[b]{\smash{\SetFigFont{10}{13.2}{rm}$\B{\phi_1 \A \psi}_{\alpha}$}}}
\end{picture}

\end{center}

Observe that $\phi_1 \wedge\psi$ and $\phi_2 \wedge\psi$ are smaller
formulas than $(\phi_1\vee\phi_2) \wedge\psi$ in the adopted
lexicographic ordering.

\item\label{os:conj} 
If $\psi,\phi_1,\phi_2$ are formulas,
then we put $\B{(\phi_1\wedge\phi_2)\wedge\psi}_\alpha$ to be
the tree with a root of degree one and the tree
$\B{\phi_1\wedge (\phi_2 \wedge\psi)}_\alpha$ 
as its subtree:

\begin{center}
\begin{picture}(0,0)%
\includegraphics{aptbezem8.pstex}%
\end{picture}%
\setlength{\unitlength}{3947sp}%
\begingroup\makeatletter\ifx\SetFigFont\undefined
\def\x#1#2#3#4#5#6#7\relax{\def\x{#1#2#3#4#5#6}}%
\expandafter\x\fmtname xxxxxx\relax \def\y{splain}%
\ifx\x\y   
\gdef\SetFigFont#1#2#3{%
  \ifnum #1<17\tiny\else \ifnum #1<20\small\else
  \ifnum #1<24\normalsize\else \ifnum #1<29\large\else
  \ifnum #1<34\Large\else \ifnum #1<41\LARGE\else
     \huge\fi\fi\fi\fi\fi\fi
  \csname #3\endcsname}%
\else
\gdef\SetFigFont#1#2#3{\begingroup
  \count@#1\relax \ifnum 25<\count@\count@25\fi
  \def\x{\endgroup\@setsize\SetFigFont{#2pt}}%
  \expandafter\x
    \csname \romannumeral\the\count@ pt\expandafter\endcsname
    \csname @\romannumeral\the\count@ pt\endcsname
  \csname #3\endcsname}%
\fi
\fi\endgroup
\begin{picture}(2615,1140)(2593,-1219)
\put(3901,-1186){\makebox(0,0)[b]{\smash{\SetFigFont{10}{13.2}{rm}$\B{\phi_1 \A (\phi_2 \A \psi)}_{\alpha}$}}}
\put(3901,-211){\makebox(0,0)[b]{\smash{\SetFigFont{10}{13.2}{rm}$(\phi_1 \A \phi_2) \A \psi, \ \alpha$}}}
\end{picture}

\end{center}

This substantiates the association of conjunctions to the right as
mentioned in Definition~\ref{def:formulas}.  Note that, again, the
definition refers to lexicographically smaller formulas.

\item\label{os:impl}
If $\psi,\phi_1,\phi_2$ are formulas,
then we put $\B{(\phi_1\ra\phi_2)\wedge\psi}_\alpha$ to be
a tree with a root of degree one. We distinguish three cases.

\begin{itemize}
\item Formula $\phi_1$ is $\alpha$-closed and $\B{\phi_1}_\alpha$
contains only failure leaves.
Then the root of\\$\B{(\phi_1 \ra\phi_2) \wedge\psi}_\alpha$ has
$\B{\psi}_\alpha$ as its subtree:
 
\begin{center}
\begin{picture}(0,0)%
\includegraphics{aptbezem14.pstex}%
\end{picture}%
\setlength{\unitlength}{3947sp}%
\begingroup\makeatletter\ifx\SetFigFont\undefined
\def\x#1#2#3#4#5#6#7\relax{\def\x{#1#2#3#4#5#6}}%
\expandafter\x\fmtname xxxxxx\relax \def\y{splain}%
\ifx\x\y   
\gdef\SetFigFont#1#2#3{%
  \ifnum #1<17\tiny\else \ifnum #1<20\small\else
  \ifnum #1<24\normalsize\else \ifnum #1<29\large\else
  \ifnum #1<34\Large\else \ifnum #1<41\LARGE\else
     \huge\fi\fi\fi\fi\fi\fi
  \csname #3\endcsname}%
\else
\gdef\SetFigFont#1#2#3{\begingroup
  \count@#1\relax \ifnum 25<\count@\count@25\fi
  \def\x{\endgroup\@setsize\SetFigFont{#2pt}}%
  \expandafter\x
    \csname \romannumeral\the\count@ pt\expandafter\endcsname
    \csname @\romannumeral\the\count@ pt\endcsname
  \csname #3\endcsname}%
\fi
\fi\endgroup
\begin{picture}(2307,1140)(2747,-1219)
\put(3901,-1186){\makebox(0,0)[b]{\smash{\SetFigFont{10}{13.2}{rm}$\B{\psi}_{\alpha}$}}}
\put(3901,-211){\makebox(0,0)[b]{\smash{\SetFigFont{10}{13.2}{rm}$(\phi_1 \ra \phi_2) \A \psi, \ \alpha$}}}
\end{picture}

\end{center}
 
\item Formula $\phi_1$ is $\alpha$-closed and
$\B{\phi_1}_\alpha$ contains at least one success leaf.
Then the root of $\B{(\phi_1 \ra\phi_2) \wedge\psi}_\alpha$ has
$\B{\phi_2 \wedge\psi}_\alpha$ as its subtree:
 
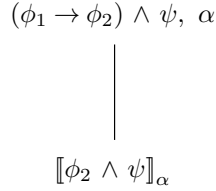
\begin{center}
\begin{picture}(0,0)%
\includegraphics{aptbezem15.pstex}%
\end{picture}%
\setlength{\unitlength}{3947sp}%
\begingroup\makeatletter\ifx\SetFigFont\undefined
\def\x#1#2#3#4#5#6#7\relax{\def\x{#1#2#3#4#5#6}}%
\expandafter\x\fmtname xxxxxx\relax \def\y{splain}%
\ifx\x\y   
\gdef\SetFigFont#1#2#3{%
  \ifnum #1<17\tiny\else \ifnum #1<20\small\else
  \ifnum #1<24\normalsize\else \ifnum #1<29\large\else
  \ifnum #1<34\Large\else \ifnum #1<41\LARGE\else
     \huge\fi\fi\fi\fi\fi\fi
  \csname #3\endcsname}%
\else
\gdef\SetFigFont#1#2#3{\begingroup
  \count@#1\relax \ifnum 25<\count@\count@25\fi
  \def\x{\endgroup\@setsize\SetFigFont{#2pt}}%
  \expandafter\x
    \csname \romannumeral\the\count@ pt\expandafter\endcsname
    \csname @\romannumeral\the\count@ pt\endcsname
  \csname #3\endcsname}%
\fi
\fi\endgroup
\begin{picture}(2307,1140)(2747,-1219)
\put(3901,-211){\makebox(0,0)[b]{\smash{\SetFigFont{10}{13.2}{rm}$(\phi_1 \ra \phi_2) \A \psi, \ \alpha$}}}
\put(3901,-1186){\makebox(0,0)[b]{\smash{\SetFigFont{10}{13.2}{rm}$\B{\phi_2 \A \psi}_{\alpha}$}}}
\end{picture}

\end{center}


\item 
In all other cases the root of $\B{(\phi_1 \ra \phi_2) \wedge\psi}_\alpha$ has
the error leaf $\err$ as its son:

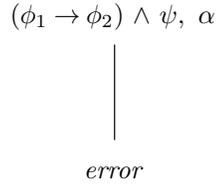
\begin{center}
\begin{picture}(0,0)%
\includegraphics{aptbezem16.pstex}%
\end{picture}%
\setlength{\unitlength}{3947sp}%
\begingroup\makeatletter\ifx\SetFigFont\undefined
\def\x#1#2#3#4#5#6#7\relax{\def\x{#1#2#3#4#5#6}}%
\expandafter\x\fmtname xxxxxx\relax \def\y{splain}%
\ifx\x\y   
\gdef\SetFigFont#1#2#3{%
  \ifnum #1<17\tiny\else \ifnum #1<20\small\else
  \ifnum #1<24\normalsize\else \ifnum #1<29\large\else
  \ifnum #1<34\Large\else \ifnum #1<41\LARGE\else
     \huge\fi\fi\fi\fi\fi\fi
  \csname #3\endcsname}%
\else
\gdef\SetFigFont#1#2#3{\begingroup
  \count@#1\relax \ifnum 25<\count@\count@25\fi
  \def\x{\endgroup\@setsize\SetFigFont{#2pt}}%
  \expandafter\x
    \csname \romannumeral\the\count@ pt\expandafter\endcsname
    \csname @\romannumeral\the\count@ pt\endcsname
  \csname #3\endcsname}%
\fi
\fi\endgroup
\begin{picture}(2307,1134)(1847,-913)
\put(3001,-886){\makebox(0,0)[b]{\smash{\SetFigFont{10}{13.2}{rm}{\em error}}}}
\put(3001, 89){\makebox(0,0)[b]{\smash{\SetFigFont{10}{13.2}{rm}$(\phi_1 \ra \phi_2) \A \psi, \ \alpha$}}}
\end{picture}

\end{center}
\end{itemize}

The above definition relies on the logical equivalence of
$\phi_1 \ra\phi_2$ and $\neg\phi_1 \vee \phi_1$,
but avoids unnecessary branching in the computation tree
that would be introduced by the disjunction.
In the Appendix, Subsection~\ref{ssec:libneg}, we explain how in the 
first case the condition that $\phi_1$ is $\alpha$-closed can be relaxed.

\item\label{os:negation} 
If $\phi,\psi$ are formulas, then to define
$\B{\neg\phi \wedge\psi}_\alpha$ we distinguish 
three cases with respect to $\phi$.
In all of them $\B{\neg\phi \wedge\psi}_\alpha$ is a tree
with a root of degree one.
\begin{itemize}
\item Formula $\phi$ is $\alpha$-closed and $\B{\phi}_\alpha$
contains only failure leaves. 
Then the root of $\B{ \neg\phi \wedge\psi}_\alpha$ has 
$\B{\psi}_\alpha$ as its subtree:

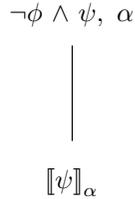
\begin{center}
\begin{picture}(0,0)%
\includegraphics{aptbezem9.pstex}%
\end{picture}%
\setlength{\unitlength}{3947sp}%
\begingroup\makeatletter\ifx\SetFigFont\undefined
\def\x#1#2#3#4#5#6#7\relax{\def\x{#1#2#3#4#5#6}}%
\expandafter\x\fmtname xxxxxx\relax \def\y{splain}%
\ifx\x\y   
\gdef\SetFigFont#1#2#3{%
  \ifnum #1<17\tiny\else \ifnum #1<20\small\else
  \ifnum #1<24\normalsize\else \ifnum #1<29\large\else
  \ifnum #1<34\Large\else \ifnum #1<41\LARGE\else
     \huge\fi\fi\fi\fi\fi\fi
  \csname #3\endcsname}%
\else
\gdef\SetFigFont#1#2#3{\begingroup
  \count@#1\relax \ifnum 25<\count@\count@25\fi
  \def\x{\endgroup\@setsize\SetFigFont{#2pt}}%
  \expandafter\x
    \csname \romannumeral\the\count@ pt\expandafter\endcsname
    \csname @\romannumeral\the\count@ pt\endcsname
  \csname #3\endcsname}%
\fi
\fi\endgroup
\begin{picture}(1693,1215)(2154,-994)
\put(3001, 89){\makebox(0,0)[b]{\smash{\SetFigFont{10}{13.2}{rm}$\neg \phi \A \psi, \ \alpha$}}}
\put(3001,-961){\makebox(0,0)[b]{\smash{\SetFigFont{10}{13.2}{rm}$\B{\psi}_\alpha$}}}
\end{picture}

\end{center}

\item Formula $\phi$ is $\alpha$-closed and $\B{\phi}_\alpha$
contains at least one success leaf.
Then the root of $\B{\neg\phi \wedge\psi}_\alpha$ has
the failure leaf $\fal$ as its son:

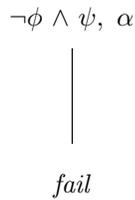
\begin{center}
\begin{picture}(0,0)%
\includegraphics{aptbezem10.pstex}%
\end{picture}%
\setlength{\unitlength}{3947sp}%
\begingroup\makeatletter\ifx\SetFigFont\undefined
\def\x#1#2#3#4#5#6#7\relax{\def\x{#1#2#3#4#5#6}}%
\expandafter\x\fmtname xxxxxx\relax \def\y{splain}%
\ifx\x\y   
\gdef\SetFigFont#1#2#3{%
  \ifnum #1<17\tiny\else \ifnum #1<20\small\else
  \ifnum #1<24\normalsize\else \ifnum #1<29\large\else
  \ifnum #1<34\Large\else \ifnum #1<41\LARGE\else
     \huge\fi\fi\fi\fi\fi\fi
  \csname #3\endcsname}%
\else
\gdef\SetFigFont#1#2#3{\begingroup
  \count@#1\relax \ifnum 25<\count@\count@25\fi
  \def\x{\endgroup\@setsize\SetFigFont{#2pt}}%
  \expandafter\x
    \csname \romannumeral\the\count@ pt\expandafter\endcsname
    \csname @\romannumeral\the\count@ pt\endcsname
  \csname #3\endcsname}%
\fi
\fi\endgroup
\begin{picture}(1693,1209)(2454,-988)
\put(3301, 89){\makebox(0,0)[b]{\smash{\SetFigFont{10}{13.2}{rm}$\neg \phi \A \psi, \ \alpha$}}}
\put(3301,-961){\makebox(0,0)[b]{\smash{\SetFigFont{10}{13.2}{rm}{\em fail}}}}
\end{picture}

\end{center}

\item 
In all other cases the root of $\B{\neg\phi \wedge\psi}_\alpha$ has
the error leaf $\err$ as its son:

\begin{center}
\begin{picture}(0,0)%
\includegraphics{aptbezem11.pstex}%
\end{picture}%
\setlength{\unitlength}{3947sp}%
\begingroup\makeatletter\ifx\SetFigFont\undefined
\def\x#1#2#3#4#5#6#7\relax{\def\x{#1#2#3#4#5#6}}%
\expandafter\x\fmtname xxxxxx\relax \def\y{splain}%
\ifx\x\y   
\gdef\SetFigFont#1#2#3{%
  \ifnum #1<17\tiny\else \ifnum #1<20\small\else
  \ifnum #1<24\normalsize\else \ifnum #1<29\large\else
  \ifnum #1<34\Large\else \ifnum #1<41\LARGE\else
     \huge\fi\fi\fi\fi\fi\fi
  \csname #3\endcsname}%
\else
\gdef\SetFigFont#1#2#3{\begingroup
  \count@#1\relax \ifnum 25<\count@\count@25\fi
  \def\x{\endgroup\@setsize\SetFigFont{#2pt}}%
  \expandafter\x
    \csname \romannumeral\the\count@ pt\expandafter\endcsname
    \csname @\romannumeral\the\count@ pt\endcsname
  \csname #3\endcsname}%
\fi
\fi\endgroup
\begin{picture}(1693,1134)(2154,-913)
\put(3001, 89){\makebox(0,0)[b]{\smash{\SetFigFont{10}{13.2}{rm}$\neg \phi \A \psi, \ \alpha$}}}
\put(3001,-886){\makebox(0,0)[b]{\smash{\SetFigFont{10}{13.2}{rm}{\em error}}}}
\end{picture}

\end{center}

There are basically two classes of formulas $\phi$ in this contingency:
those that are not $\alpha$-closed and those for which
$\B{\phi}_\alpha$ contains no success leaf and
at least one error leaf. In Subsection~\ref{ssec:libneg}
we give some examples
of formulas in the first class and show how in some special
cases their negation can still be evaluated in a sound way.
\end{itemize}
\item\label{os:exquant} The case of $\exists x\;\phi \wedge\psi$
requires the usual care with bound variables to avoid name clashes.
Let $\alpha$ be a valuation. First, we require
that the variable $x$ does not occur in the domain of $\alpha$.
Second, we require that the variable $x$ does not occur in $\psi$.
Both requirements are summarized by phrasing that $x$ is \emph{fresh}
with respect to $\alpha$ and $\psi$. They can be met by
appropriately renaming the bound variable $x$.
 
With $x$ fresh as above we define $\B{\exists x\;\phi \wedge\psi}_\alpha$
to be the tree with a root of degree one and 
$\B{\phi\wedge\psi}_\alpha$ as its subtree:

\begin{center}
\begin{picture}(0,0)%
\includegraphics{aptbezem12.pstex}%
\end{picture}%
\setlength{\unitlength}{3947sp}%
\begingroup\makeatletter\ifx\SetFigFont\undefined
\def\x#1#2#3#4#5#6#7\relax{\def\x{#1#2#3#4#5#6}}%
\expandafter\x\fmtname xxxxxx\relax \def\y{splain}%
\ifx\x\y   
\gdef\SetFigFont#1#2#3{%
  \ifnum #1<17\tiny\else \ifnum #1<20\small\else
  \ifnum #1<24\normalsize\else \ifnum #1<29\large\else
  \ifnum #1<34\Large\else \ifnum #1<41\LARGE\else
     \huge\fi\fi\fi\fi\fi\fi
  \csname #3\endcsname}%
\else
\gdef\SetFigFont#1#2#3{\begingroup
  \count@#1\relax \ifnum 25<\count@\count@25\fi
  \def\x{\endgroup\@setsize\SetFigFont{#2pt}}%
  \expandafter\x
    \csname \romannumeral\the\count@ pt\expandafter\endcsname
    \csname @\romannumeral\the\count@ pt\endcsname
  \csname #3\endcsname}%
\fi
\fi\endgroup
\begin{picture}(1847,1140)(2977,-1219)
\put(3901,-1186){\makebox(0,0)[b]{\smash{\SetFigFont{10}{13.2}{rm}$\B{\phi \A \psi}_{\alpha}$}}}
\put(3901,-211){\makebox(0,0)[b]{\smash{\SetFigFont{10}{13.2}{rm}$(\te x \phi) \A \psi, \ \alpha$}}}
\end{picture}

\end{center}

Thus the operational semantics of $\exists x\;\phi \wedge\psi$ is,
apart from the root of degree one, identical to that of $\phi\wedge\psi$.
This should not come as a surprise, as $\exists x\;\phi \wedge\psi$
is logically equivalent to $\exists x\;(\phi\wedge\psi)$ when $x$
does not occur in $\psi$.

Observe that success leaves of $\B{\phi\wedge\psi}_\alpha$,
and hence of $\B{\exists x\;\phi \wedge\psi}_\alpha$,
may or may not contain an assignment for $x$. For example,
$\exists x\;x=3 \wedge\psi$ yields an assignment for $x$,
but $\exists x\;3=3 \wedge\psi$ does not. In any case the assignment
for $x$ is not relevant for the formula as a whole, 
as the bound variable $x$ is assumed to be fresh.
In an alternative approach, the possible assignment for $x$ could be deleted.
\HB
\end{enumerate}
\end{definition}

To apply the above computation mechanism to arbitrary first-order
formulas we first replace all occurrences of a universal quantifier
$\fa$ by $\neg \te \neg$ and rename the bound variables so that no
variable appears in a formula both bound and free.

Further, to minimize the possibility of generating errors it is useful
to delete occurrences of double negations, that is, to replace every
subformula of the form $\neg \neg \psi$ by $\psi$.

\section{Soundness and Completeness}
\label{sec:souncom}

The computation mechanism defined in the previous section
attempts to find a valuation that makes the original
formula true if this formula is satisfiable, and otherwise it 
reports a failure. The lexicographic ordering used in 
Definition~\ref{def:opersem} guarantees that for any formula
the computation tree is finite.
In this section we prove correctness 
and completeness of this mechanism.

We start with an easy lemma which is helpful to keep track of valuations,
followed by a definition.

\begin{lemma}\label{lem:extval} For every formula $\phi$ and valuation $\alpha$,
$\B{\phi}_\alpha$ contains only valuations extending $\alpha$
with pairs $x/d$, where $x$ occurs free in $\phi$ or
appears existentially quantified in $\phi$.
Moreover, if $\phi$ is $\alpha$-closed then $\B{\phi}_\alpha$ 
contains only valuations extending $\alpha$ with variables
that appear existentially quantified in $\phi$.
\end{lemma}
\Proof
By induction on the lexicographic ordering of formulas
as given in Definition~\ref{def:tree}.\HB
 
\begin{definition}[status of computation tree]\label{def:treestat}
A computation tree is
\begin{itemize}
\item {\em successful\/} if it contains a success leaf,
\item {\em failed\/} if it contains only failure leaves,
\item {\em determined\/} if it is either successful or failed,
that is, it either contains a success leaf or contains only
failure leaves. 
\HB
\end{itemize}
\end{definition}
Note that according to this definition a successful tree can contain
the $\err$ leaves.  This means that the $\err$ leaves differ from
Prolog's run-time errors.  In fact, in a top-down implementation of
the proposed computation mechanism the depth-first search traversal of
a computation tree should {\em not} abort but rather backtrack upon
encounter of such a leaf and continue, if possible, in a search for a
successful leaf.
 
We can now state the desired correctness result.

\begin{theorem}[Soundness]\label{thm:soundn}
Let $\phi$ be a formula and $\alpha$ a valuation.

\begin{enumerate}\smallromani
\item 
If $\B{\phi}_\alpha$ contains a success leaf labelled with $\alpha'$,
then $\alpha'$ extends $\alpha$ and $\fa(\phi^{\alpha'})$ is true.
(In particular $\te(\phi^{\alpha})$ is true in this case.)
\item If  $\B{\phi}_\alpha$ is failed, then
$\te(\phi^{\alpha})$ is false. 
\end{enumerate}
\end{theorem}

\Proof
See Appendix, Subsection~\ref{ssec:soundn}.
\HB
\VV

The computation mechanism defined in Section~\ref{sec:souncom} is
obviously incomplete due to the possibility of errors.  The following
results states that, in the absence of errors, this mechanism is
complete. 
\begin{theorem}[Restricted Completeness] \label{thm:completeness}
Let $\phi$ be a formula and $\alpha$ a valuation such that
$\B{\phi}_\alpha$ is determined.

\begin{enumerate}\smallromani
\item Suppose that $\te(\phi^{\alpha})$ is true. 
Then the tree $\B{\phi}_\alpha$ is successful.

\item Suppose that $\te(\phi^{\alpha})$ is false. 
Then the tree $\B{\phi}_\alpha$ is failed.

\end{enumerate}
\end{theorem}
\Proof
See Appendix, Subsection~\ref{ssec:completeness}.
\HB

\VV

Admittedly, this result is very weak in the sense that any computation
mechanism that satisfies the above soundness theorem 
also satisfies the restricted completeness theorem.

It is useful to point out that the computation mechanism of 
Section~\ref{sec:compu} used in the above theorems is by no means a simple
counterpart of the provability relation of the first-order logic.

For the sake of further discussion let us say that 
two formulas $\phi$ and
$\psi$ are {\em equivalent\/}
if 
\begin{itemize}
\item 
the computation
tree $\B{\phi}_\empval$
is successful iff the computation
tree $\B{\psi}_\empval$ is successful and in that case
both computation trees have the same set of successful leaves,

\item $\B{\phi}_\empval$ is failed iff $\B{\psi}_\empval$ is failed.
\end{itemize}

Then
$\phi \A \psi$ is not equivalent to $\psi \A \phi$
(consider ${x=0}\wedge{x<1}$ and ${x<1}\wedge{x=0}$)
and $\neg (\phi \A \psi)$ is not equivalent to 
$\neg \phi \Or \neg \psi$
(consider $\neg({x=0}\wedge{x=1})$ and $\neg({x=0})\vee\neg({x=1})$.
In contrast, 
$\phi \Or \psi$ {\em is\/} equivalent to $\psi \Or \phi$.

We can summarize this treatment of the connectives by saying that we
use a sequential conjunction and a parallel disjunction. The above
notion of equivalence deviates from the usual one, for example
de Morgan's Law is not valid.

A complete axiomatization of the equivalence relation induced by the
computation mechanism of Section~\ref{sec:compu} is an interesting
research topic.

\section{Extensions}
\label{sec:extensions}

The language defined up to now is clearly too limited as a formalism
for programming. Therefore we discuss a number of 
extensions of it that are convenient for programming purposes.
These are: non-recursive procedures, sorts (i.e., types),
arrays and bounded quantification.

\subsection{Non-recursive Procedures}
\label{subsec: nonrec}

We consider here non-recursive procedures. These can easily be 
introduced in our framework using the well-known
{\em extension by definition\/} mechanism 
(see, e.g., \citeasnoun{Sho67}[pages 57-58]).

More specifically, 
consider a first-order formula $\psi$ with the free variables
$x_1, \LL, x_n$. Let $p$ be a {\em new\/} $n$-ary relation symbol.
Consider now the formula
\[
p(x_1, \LL, x_n) \lra \psi
\]
that we call the {\em definition\/} of $p$.

Suppose that, by iterating the above procedure, we have a collection $P$
of definitions of relation symbols. We assume furthermore that
the fixed but arbitrary interpretation has been extended with
interpretations of the new relation symbols in such a way
that all definitions in $P$ become true. There is only one
such extension for every initial interpretation.

Let $\phi$ be a formula in the extended first-order language,
that is, with atoms $p(t_1, \LL, t_n)$ from $P$ included.
We extend the computation mechanism $\B{\phi}_\alpha$ 
of Section~\ref{sec:compu}, by adding at the beginning of 
Clause~\ref{os:atom} in Definition~\ref{def:tree} the
following item for handling atoms $p(t_1, \LL, t_n)$ from $P$.

\begin{itemize}

\item Atom $A$ is of the form $p(t_1, \LL, t_n)$, where
$p$ is a defined relation symbol with the definition
\[
p(x_1, \LL, x_n) \lra \psi_p.
\]
Then the root of $\B{A\wedge\psi}_\alpha$ has $\B{\psi_{p}\C{x_1/t_1, \LL,
x_n/t_n} \A \psi}_\alpha$ as its subtree:

\begin{center}
\begin{picture}(0,0)%
\includegraphics{aptbezem26.pstex}%
\end{picture}%
\setlength{\unitlength}{3947sp}%
\begingroup\makeatletter\ifx\SetFigFont\undefined
\def\x#1#2#3#4#5#6#7\relax{\def\x{#1#2#3#4#5#6}}%
\expandafter\x\fmtname xxxxxx\relax \def\y{splain}%
\ifx\x\y   
\gdef\SetFigFont#1#2#3{%
  \ifnum #1<17\tiny\else \ifnum #1<20\small\else
  \ifnum #1<24\normalsize\else \ifnum #1<29\large\else
  \ifnum #1<34\Large\else \ifnum #1<41\LARGE\else
     \huge\fi\fi\fi\fi\fi\fi
  \csname #3\endcsname}%
\else
\gdef\SetFigFont#1#2#3{\begingroup
  \count@#1\relax \ifnum 25<\count@\count@25\fi
  \def\x{\endgroup\@setsize\SetFigFont{#2pt}}%
  \expandafter\x
    \csname \romannumeral\the\count@ pt\expandafter\endcsname
    \csname @\romannumeral\the\count@ pt\endcsname
  \csname #3\endcsname}%
\fi
\fi\endgroup
\begin{picture}(3523,1140)(1539,-919)
\put(3301, 89){\makebox(0,0)[b]{\smash{\SetFigFont{10}{13.2}{rm}$A \A \psi, \ \alpha$}}}
\put(3301,-886){\makebox(0,0)[b]{\smash{\SetFigFont{10}{13.2}{rm}$\B{\psi_{p}\C{x_1/t_1, \LL, x_n/t_n} \A \psi}_\alpha$}}}
\end{picture}

\end{center}
Here $\psi_{p}\C{x_1/t_1, \LL, x_n/t_n}$ stands for the result of
substituting in $\psi_{p}$ the free occurrences of the variables $x_1,
\LL, x_n$ by  $t_1, \LL, t_n$, respectively.
\end{itemize}

The proof of the termination of this extension of the computation mechanism
introduced in Section~\ref{sec:compu} relies on a refinement of the
lexicographic ordering used in Definition~\ref{def:tree},
taking into account the new atoms.

The above way of handling defined relation symbols obviously
corresponds to the usual treatment of procedure calls in
programming languages.

The soundness and completeness results can easily be extended
to the case of declared relation symbols. In this version
truth and falsity refer to the extended interpretation.

So far for \emph{non-recursive} procedures.

\subsection{Sorts}
 
In this subsection we introduce sorts (i.e., types).
The extension of one-sorted to many-sorted first-order logic is
standard.  It requires a refinement of the notion of signature:
arities are no longer just numbers, but have to specify the sorts of
the arguments of the function and predicate symbols, as well as the sorts
of the function values. Terms and atoms are 
well-formed if the sorts of the arguments comply with the signature.
In quantifying a variable, its sort should be made explicit (or should
at least be clear from the context). 

Interpretations for many-sorted first-order languages are obtained
by assigning to each sort a non-empty domain and by assigning to each
function symbol and each predicate symbol respectively an
appropriate function and relation on these sorts.

Sorts can be used to model
various basic data types occurring in programming practice:
integers, booleans, characters, but also compound data types
such as arrays.

\subsection{Arrays}

Arrays can be modelled as vectors or matrices,
using projection functions that are given a \emph{standard interpretation}.
Given a sort for the indices (typically, a segment of integers or
a product of segments) and a sort for the elements of the array,
we add a sort for arrays of the corresponding type to the signature.
We also add to the language {\em array variables\/}, 
or {\em arrays\/} for short,
to be interpreted as arrays in the standard interpretation.

We use the letters $a,b,c$ to denote arrays and 
to distinguish arrays from objects of other sorts.
We write $a[t_1, \LL, t_n]$ to denote the projection of the array
$a$ on the index $[t_1, \LL, t_n]$,
akin to the use of subscripted variables in programming languages.
The standard interpretation of each projection function maps
a given array and a given index to the correct element.
Thus subscripted variables are simply terms.
These terms are handled by means 
of an extension of the computation mechanism of Section~\ref{sec:compu}.

A typical example of the use of such a term is 
the formula $a[0,0]=1$, which should be matched with the formula $x=1$
in the sense that the evaluation of each equality can result in an
assignment of the value 1 to a variable, either  $a[0,0]$ or $x$. So
we view $a[0,0]$ as a variable and not as a compound term.

To this end we extend a number of notions introduced in 
the previous section.

\begin{definition}
An {\em array valuation\/} is a finite mapping whose elements
are of the form $a[d_1, \LL, d_n]/d$, where $a$ is an $n$-ary array
symbol and $d_1, \LL, d_n, d$ are domain elements.
An {\em extended valuation\/} is a finite mapping that is a union
of a valuation and an array valuation.
\HB
\end{definition}

The idea is that an element $a[d_1, \LL, d_n]/d$ of an array
valuation assigns the value $d$ to the (interpretation of)
array $a$ applied to the arguments $d_1, \LL, d_n$. 
Then, if the terms $t_1, \LL, t_n$ evaluate to the domain
elements $d_1, \LL, d_n$ respectively, 
the term $a[t_1, \LL, t_n]$ evaluates to $d$.
This simple inductive clause yields an extension of the notion 
of evaluation $t^{\alpha}$, where $\alpha$ is an extended valuation, 
to terms $t$ in the presence of arrays.
The notions of an $\alpha$-closed term and an $\alpha$-assignment
are now somewhat more complicated to define.

\begin{definition}
Consider an extended valuation $\alpha$.
\begin{itemize}

\item A variable $x$ is {\em $\alpha$-closed\/} if
for some $d$ the pair $x/d$ is an element of $\alpha$.

\item A term  $f(t_1, \LL, t_n)$, with $f$ a function symbol,
is {\em $\alpha$-closed\/} if each term $t_i$ is $\alpha$-closed.

\item A term $a[t_1, \LL, t_n]$ is {\em $\alpha$-closed\/} if
each term $t_i$ is $\alpha$-closed and evaluates to a domain element $d_i$
such that for some $d$ the pair $a[d_1, \LL, d_n]/d$ is an element of 
$\alpha$.

\end{itemize}

An equation $s = t$ is an {\em $\alpha$-assignment\/} if either

\begin{itemize}

\item one side of it, say $s$, is a variable that is not $\alpha$-closed
and the other, $t$, is an $\alpha$-closed term, or

\item one side of it, say $s$, is of the form $a[t_1, \LL, t_n]$, where 
each $t_i$ is $\alpha$-closed but $a[t_1, \LL, t_n]$ is not $\alpha$-closed,
and the other, $t$, is an $\alpha$-closed term.\HB
\end{itemize}
\end{definition}

The idea is that an array $a$ can be assigned a value at a selected position by
evaluating an $\alpha$-assignment $a[t_1, \LL, t_n] = t$. Assuming the 
terms $t_1, \LL, t_n, t$ are $\alpha$-closed and evaluate respectively to
$d_1, \LL, d_n, d$, the evaluation of $a[t_1, \LL, t_n] = t$
results in assigning the value $d$ to the array $a$ at the
position $d_1, \LL, d_n$. 

With this extension of the notions of valuation and $\alpha$-assignment
we can now apply the computation mechanism of Section~\ref{sec:compu}
to first-order formulas with arrays. The corresponding extensions
of the soundness and completeness theorems of
Section~\ref{sec:souncom} remain valid.

\subsection{Bounded quantification}

In this subsection we show how to extend the language with a form of
bounded quantification that essentially amounts to the
generalized conjunction
and disjunction. We treat bounded quantification with
respect to the integer numbers, but the approach can easily
be generalized to data types with the same discrete and ordered
structure as the integers. 

\begin{definition}[bounded quantification]\label{def:bquant}
Let $\alpha$ be a valuation and let $\phi(x)$ be a formula
with $x$ not occurring in the domain of $\alpha$.
Furthermore, let $s,t$ be terms of integer type.
We assume the set of formulas to be extended in such a way that
also $\te x\in[s..t]\;\phi(x)$ and $\fa x\in[s..t]\;\phi(x)$
are formulas. The computation trees of these formulas have a
root of degree one and depend on $s$ and $t$
in the following way.
\begin{itemize}
\item If $s$ or $t$ is not $\alpha$-closed, then the roots of both
$\B{\te x\in[s..t]\;\phi(x)}_\alpha$ and
$\B{\fa x\in[s..t]\;\phi(x)}_\alpha$ have the error leaf $\err$
as its son.
\item If $s$ and $t$ are $\alpha$-closed and $s^\alpha > t^\alpha$, 
then the root of $\B{\te x\in[s..t]\;\phi(x)}_\alpha$ has
the failure leaf $\fal$ as its son and the root of
$\B{\fa x\in[s..t]\;\phi(x)}_\alpha$ has a success leaf $\alpha$
as its son.
\item If $s$ and $t$ are $\alpha$-closed and $s^\alpha \leq t^\alpha$,
then 
\begin{itemize}
\item[-] the root of $\B{\te x\in[s..t]\;\phi(x)}_\alpha$ has 
$\B{\phi(x)\vee\te y\in[s{+}1..t]\;\phi(y)}_{\alpha\cup\{x/{s^\alpha}\}}$ 
as its subtree,
\item[-] the root of $\B{\fa x\in[s..t]\;\phi(x)}_\alpha$ has
$\B{\phi(x)\wedge\fa y\in[s{+}1..t]\;\phi(y)}_{\alpha\cup\{x/{s^\alpha}\}}$ 
as its subtree.
\end{itemize}
In both cases $y$ should be a fresh variable with respect to $\alpha,\phi(x)$
in order to avoid name clashes.
\end{itemize}
The soundness and completeness results can easily be extended
to include bounded quantification.
\HB
\end{definition}

\section{Relation to Other Approaches}
\label{sec:related}

The work here discussed is related in many interesting ways to a number of
seminal papers on logic, logic programming and constraint logic programming.

\subsection{Definition of Truth compared to Formulas as Programs}

First, it is instructive to compare our approach to the inductive
definition of truth given in \citeasnoun{Tar33}. This definition can
be interpreted as an algorithm that, given a first-order language $L$,
takes as input an interpretation $I$ of $L$ and a formula $\phi$ of
$L$, and yields as output the answer to the question whether the
universal closure of $\phi$ is true in $I$.  This algorithm is not
effective because of the way quantifiers are dealt with. This is unavoidable
since truth is undecidable for many languages and interpretations, for instance
Peano arithmetic and its standard model.

In the formulas as programs approach the initial problem is modified 
in that one asks for a constructive answer to the question whether a formula 
is satisfiable in an interpretation. The algorithm proposed here is
effective at the cost of occasionally terminating abnormally in an error.

\subsection{Relation to Logic Programming}

Some forty years later, in his seminal paper \citeasnoun{Kow74}
proposed to use first-order logic as a computation formalism. This led
to logic programming.  However, in spite of the paper's title, only a
subset of first-order logic is used in his proposal, namely the one
consisting of Horn clauses.  This restriction was essential since
what is now called SLD-resolution was used as the computation mechanism.

In the discussion we first concentrate on the syntax matters and then
focus on the computation mechanism.

The restriction of logic programs and goals to Horn clauses was
gradually lifted in \citeasnoun{Cla78}, by allowing negative literals
in the goals and in clause bodies, in \citeasnoun{LT84}, by allowing
arbitrary first-order formulas as goals and clause bodies, and in
\citeasnoun{LMR92} by allowing disjunctions in the clause heads.  In
each case the computation mechanism of SLD-resolution was suitably
extended, either by introducing the negation as failure rule, or by
means of transformation rules, or by generalizing so-called linear
resolution.

   From the syntactic point of view
our approach is related to that of
\citeasnoun{LT84}.  Appropriate transformation rules are used there
to get rid of quantifiers, disjunctions and the applications of
negation to non-atomic formulas. So these features of first-order
logic are interpreted in an indirect way.  It is useful to point out
that the approach of \citeasnoun{LT84} was implemented in the
programming language G\"{o}del of \citeasnoun{HL94}.

Further, it should be noted that bounded quantifiers and arrays
were also studied in logic programming. In particular, they are used
in the specification language Spill of \citeasnoun{KM97} that allows
us to write executable, typed, specifications in the logic programming
style.  Other related references are \citeasnoun{Vor92},
\citeasnoun{BB93} and \citeasnoun{Apt96}.

So from the syntactic point of view our approach does not seem to
differ from logic programming in an essential way. The difference
becomes more apparent when we analyze in more detail the underlying computation
mechanism.

To this end it is useful to recall that in logic programming the
computing process takes place implicitly over the free algebra of all
terms and the values are assigned to variables by means of
unification.  The first aspect can be modelled in the formulas as
programs approach by choosing a {\em term interpretation\/}, so an
interpretation the domain $D$ of which consists of all terms and such
that each $n$-ary function symbol $f$ is mapped to a function $f_D$
that assigns to elements (so terms) $t_1, \LL, t_n$ of $D$ the term
$f(t_1, \LL, t_n)$. With this choice our use of $\alpha$-assignment
boils down to an instance of matching which in turn is a special case
of unification.

Unification in logic programming can be more clearly related to
equality by means of the so-called homogenization process the purpose
of which is to remove non-variable terms from the clauses heads. For
instance,

{\tt append(x1,ys,z1) <- x1=[x|xs], z1=[x|zs], append(xs,ys,zs)}

\NI
is a homogenized form of the more compact clause

{\tt append([x|xs],ys,[x|zs]) <- append(xs,ys,zs)}.

\NI
To interpret the equality in the right way the single clause

{\tt x = x <-} 

\NI
should then be added. This enforces the ``is unifiable with''
interpretation of equality. So the homogenization process reveals 
that logic programming relies on a more general interpretation of equality 
than the formulas as programs approach. It allows one to avoid generation 
of errors for all equality atoms.

In conclusion, from the computational point of view,
the logic programming approach is at the same
time a restriction of the formulas as programs approach to the term
interpretations and a generalization of this approach in which all equality
atoms can be safely evaluated.

\subsection{Relation to Pure Prolog}

By pure Prolog we mean here a subset of Prolog formed by the programs and goals
that are Horn clauses.

Programming in Prolog and in its pure subset relies heavily on
lists and recursion.  As a result termination is one of the crucial
issues. This led to an extensive study of methods that allow us to
prove termination of logic and Prolog programs (see \citeasnoun{DD94}
for a survey of various approaches).

In contrast, our approach to programming 
is based on arrays and iteration that is realized by means of bounded 
quantification. These constructs are guaranteed to terminate. In
fact, it is striking how far one can go in programming in this style
without using recursion. If the reader is not convinced by the
example given of Section~\ref{sec:squares} below, he/she is invited to
consult other examples in \citeasnoun{Vor92}
and \citeasnoun{ABPS98a}.

In the formulas as programs approach the absence of recursion makes it
possible to analyze queries without explicit presence of procedures,
by systematically replacing procedures by their bodies.  This allows
us to represent each program as a single query and then rely on the
well-understood Tarskian semantics of first-order logic.

In the standard logic programming setting very few interesting
programs can be represented in this way. In fact, as soon as recursion
is used, a query has to be studied in the context of a program that 
defines the recursive procedures. As
soon as negation is also present, a plethora of different semantics
arises --- see e.g. \citeasnoun{AB94}.
Finally, in the presence of recursion
it is difficult to account for Prolog's selection rule
in purely semantic terms.

\subsection{Relation to Pure Prolog with Arithmetic}

By pure Prolog with arithmetic we mean here an extension of pure
Prolog by features that support arithmetic, so Prolog's arithmetic
relations such as ``=:='' and the Prolog evaluator operator {\tt is}.

These features allow us to compute in the presence of arithmetic but
in a clumsy way as witnessed by the example of
formula (\ref{eq:no-prolog}) of Subsection~\ref{ssec:informal}
and its elaborated representation in Prolog
in Subsection~\ref{ssec:rationale}.

Additionally, a possibility of abnormal termination in an error arises.
Indeed, both  arithmetic relations and the {\tt is} operator 
introduce a possibility of run-time errors,
a phenomenon absent in pure Prolog. For instance, the query
{\tt X is Y} yields an error and so does {\tt X =:= Y}.

In contrast, in the formulas as programs approach arithmetic
can be simply modelled by adding the sorts of integers and of reals.
The $\alpha$-assignment then deals correctly with arithmetic
expressions because it relies on automatic evaluation of terms. 
This yields a simpler and more uniform approach to arithmetic in which no 
new special relation symbols are needed.

\subsection{Relation to Constraint Logic Programming}
\label{subsec:clp}
The abovementioned deficiencies of pure Prolog with arithmetic have
been overcome in constraint logic programming, an approach to
computing that generalizes logic programming.  In what follows we
concentrate on a specific approach, the generic scheme CLP(X) of
\citeasnoun{jaffar-constraint-87} that generalizes pure Prolog by
allowing constraints.  In this scheme atoms are divided into those
defined by means of clauses and those interpreted in a direct way. The
latter ones are called constraints.

In CLP(X), as in our case, the computation is carried out over an
arbitrary interpretation. At each step (instead of the
unification test of logic programming and its application if
it succeeds) satisfiability of the so far encountered constraints is
tested. A computation is successful if the last query 
consists of constraints only.

There are two differences between the formulas as programs approach
and the CLP(X) scheme.  The first one has to do with the fact that in
our approach full first-order logic is allowed, while in the latter --- as
in logic programming and pure Prolog --- Horn clauses are used.

The second one concerns the way values are assigned. 
In our case the only way to assign values to variables is
by means of an $\alpha$-assignment, while in the  CLP(X) scheme
satisfiability of constraints guides the computation
and output is identified with a set of constraints (that still have to
be solved or normalized).

The CLP(X) approach to computing has been realized in a number of
constraint logic programming languages, notably in the CLP(${\cal R}$)
system of \citeasnoun{jaffar-clpr} that is an instance of the CLP(X)
scheme with a two-sorted structure that consists of reals and terms.
In this system formula (\ref{eq:no-prolog}) of Subsection
\ref{ssec:informal} can be directly run as a query.

Once negation is added to the CLP(X) scheme (it is in fact present in
CLP(${\cal R}$)), the extension of the CLP(X) syntax to full
first-order logic could be achieved by using the approach
\citeasnoun{LT84} or by extending the computation mechanism along the
lines of Section~\ref{sec:compu}.

So, ignoring the use of the first-order logic syntax in the formulas
as programs approach and the absence of (recursive) procedures that
could be added to it, the main difference between this approach and the
CLP(X) scheme has to do with the fact that in the former only very
limited constraints are admitted, namely ground atoms and
$\alpha$-assignments. In fact, these are the only constraints that can
be resolved directly.

So from this point of view the formulas as programs approach is less
general than constraint logic programming, as embodied in the CLP(X)
scheme.  However, this more limited approach does not rely on the
satisfiability procedure for constraints (i.e., selected atomic formulas),
or any of its approximations used in specific implementations. 
In fact, the formulas as programs approach
attempts to clarify how far constraint logic programming approach can
be used without any reliance on external procedures that deal with
constraint solving or satisfiability.

\subsection{Formulas as Programs versus Formulas as Types}
 
\def\lef{{\mathit left}}
\def\rig{{\mathit right}}
\def\exi{{\mathit ex}}
 
In the so-called \emph{formulas as types} approach, also called
the Curry-Howard-De Bruijn interpretation (see e.g. \citeasnoun{TD})
(constructive) proofs of
a formula are terms whose type is the formula in question.
The type corresponding to a formula
can thus be viewed as the (possibly empty) set of all proofs
of the formula. Here `proof' refers to an operational notion of proof,
in which
\begin{itemize}
\item a proof of $\phi \vee \psi$ is either $\lef(p)$ with $p$
a proof of $\phi$, or $\rig(p)$ with $p$ a proof of $\psi$;
\item a proof of $\phi \wedge \psi$ is a pair $\langle p,q\rangle$
consisting of a proof $p$ of $\phi$ and and a proof $q$ of $\psi$;
\item  a proof of an implication $\phi \ra \psi$
is a function that maps proofs of $\phi$ to proofs of $\psi$;
\item a proof of $\forall x~\phi(x)$ is a function that maps
domain elements $d$ to proofs of $\phi(d)$;
\item a proof of $\exists x~\phi(x)$ is of the form $\exi(d,p)$
with domain element $d$ a witness for the existential statement,
and $p$ a proof of $\phi(d)$.
\end{itemize}
 
Such proofs can be taken as programs. For example,
a constructive proof of $\forall x~\exists y~\phi(x,y)$ is a function 
that maps
$d$ to an expression of the form $\exi(e_d,p_d)$ with
$p_d$ a proof of $\phi(d,e_d)$. After extraction of the witness $e_d$
the proof yields a program computing $e_d$ from $d$.
 
The main difference between formulas as types and formulas as
programs is that in the latter approach not the proofs
of the formulas, but the formulas themselves have an operational
interpretation. To illustrate this difference, consider
the computation tree of formula (\ref{eq:no-prolog})
in Fig.~\ref{fig:ctree} with its proof:
$$
\exi(3,\exi(2,
\langle\rig(p_{3=3}),\langle\rig(p_{2=2}),p_{2*3=3*2}\rangle\rangle
))
$$
Here $p_A$ is a proof of $A$, for each true closed atom $A$.
 
Observe that in the above proof the witnesses $3$ and $2$ for $x$ and
$y$, respectively, \emph{have to be given beforehand}, whereas in
our approach they are computed.  In the formulas as programs
approach the proofs
are constructed in the successful branches of the computation tree and
the computation is guided by the search for such a proof. Apart from
differences in syntax, the reader will recognize the above proof in
the successful branch of Figure~\ref{fig:ctree}.
 
Given the undecidability of the first-order logic, there is a price to be
paid for formulas programs. It consists of the possibility of abnormal
termination in an error.

\section{\almazero{}}
\label{sec:alma0}

We hope to have convinced the reader that the formulas as programs
approach, though closely related to logic programming, differs from
it in a number of crucial aspects.

This approach to programming has been realized in the implemented
programming language \almazero{} \cite{ABPS98a}. 
A similar approach to programming has been taken in the 2LP language
of \citeasnoun{MT95b}. 2LP (which stands for ``logic programming and linear
programming") uses C syntax and has been designed for constraint
programming in the area of optimization.

\almazero{} is an
extension of a subset of Modula-2 that includes nine new features
inspired by the logic programming paradigm. We briefly recall those
that are used in the sequel and refer to
\citeasnoun{ABPS98a} for a detailed presentation.

\begin{itemize}
\item Boolean expressions can be used as statements and vice versa.
  A boolean expression that is used as a statement and evaluates to
  {\tt FALSE} is identified with a {\em failure}.

\item {\em Choice points} can be created by the non-deterministic
  statements {\tt ORELSE} and {\tt SOME}. The former is a dual of the 
  statement composition and the latter is a dual of the {\tt FOR}
  statement.  Upon failure the control returns to the most recent
  choice point, possibly within a procedure body, and the computation
resumes with the next branch in the state in which the previous branch
was entered.

\item The notion of {\em initialized} variable is introduced and the
  equality test is generalized to an assignment statement in case one
  side is an uninitialized variable and the other side an
  expression with known value.

\item A new parameter passing mechanism, {\em call by mixed
    form}, denoted by the keyword {\tt MIX}, is introduced for
variables of simple type. It works as
  follows: If the actual parameter is a variable, then it is passed by
  variable.  If the actual parameter is an expression that is not a
  variable, its value is computed and assigned to a new variable $v$
  (generated by the compiler): it is $v$ that is then passed by
  variable.  So in this case the call by mixed form boils down to call
  by value.  Using this parameter mechanism we can pass both expressions with
known values and uninitialized variables as actual parameters. This
makes it possible to use a single procedure both for testing and
computing.
\end{itemize}

For efficiency reasons the \almazero{} implementation does not
realize faithfully the computation mechanism of
Section~\ref{sec:compu} as far as the errors are concerned.  First, an
evaluation of an atom that is not $\alpha$-closed and is not an
$\alpha$-assignment yields a run-time error. On the other hand, in the
other two cases when the evaluation ends with the $\err$ leaf, in the
statements {\tt NOT S} and {\tt IF S THEN T END}, the computation
process of \almazero{} simply proceeds.

The rationale for this decision is that the use of insufficiently
instantiated atoms in \almazero{} programs is to be discouraged
whereas the catching of other two cases for errors would be
computationally prohibitive.
In this respect the implementation of \almazero{} follows
the same compromise as the implementations of Prolog.

We now associate with each first-order formula $\phi$ an
\almazero{} statement ${\cal T}(\phi)$. This is done by induction
on the structure of the formula $\phi$. The translation process is
given in Table~\ref{table: table1}.

\begin{table}[htdp]
\begin{center}  
\begin{tabular}{|l|l|} \hline

Formula                     &   \almazero{} construct      \\ \hline
$A$   (atom)                &   $A$ \\
$\phi_1 \Or  \phi_2$        &   {\tt EITHER} ${\cal T}(\phi_1)$ {\tt ORELSE} ${\cal T}(\phi_2)$ {\tt END}\\
$\phi_1 \A   \phi_2$        &   ${\cal T}(\phi_1); {\cal T}(\phi_2)$        \\
$\phi \ra   \psi$           &   {\tt IF}  ${\cal T}(\phi)$ {\tt THEN}  ${\cal T}(\psi)$ {\tt END} \\
$\neg \phi$                 &   {\tt NOT} ${\cal T}(\phi)$  \\
$\te x \phi(x, \bar{y})$    &   {\tt p}$(\bar{y})$, where the procedure {\tt p} is defined by \\
                            &   {\tt PROCEDURE p(MIX $\bar{y}: \bar{\tt T});$} \\
                            &   {\tt VAR} {\em x}~:~{\tt T;} \\
                            &   {\tt BEGIN} \\
                            &   \quad ${\cal T}(\phi(x, \bar{y}))$ \\
                            &   {\tt END;} \\
                            &   where {\tt T} is the type (sort) of the variable $x$ and  \\
                            & $\bar{\tt T}$ is the sequence of types of the variables in $\bar{y}$. \\
$\te x \in [s..t] \phi$     & {\tt SOME} $x := s$ {\tt TO} $t$ {\tt DO} $ {\cal T}(\phi)$ {\tt END}  \\
$\fa x \in [s..t] \phi$     & {\tt FOR} $x := s$ {\tt TO} $t$ {\tt DO} $ {\cal T}(\phi)$ {\tt END}  \\
\hline
\end{tabular}
\caption{Translation of formulas into \almazero{} statements. 
\label{table: table1}}
 
\end{center}
\end{table}

This translation allows us to use in the sequel \almazero{} syntax
to present specific formulas.

\section{Example: Partitioning a Rectangle into Squares}
\label{sec:squares}

\begin{figure}[htpb]
{\scriptsize
$$\xymatrix@R=0.7em@C1.18em{
&1&&&&nx&&
&1&&&&nx&\\
1&4\Line[r]\Line[d]&4\Line[r]&4\Line[r]&\Line[ddd]\Linedots[rr]&&
\Linedots[dddddd]&
 &4\Line[r]\Line[d]&4\Line[r]&4\Line[r]&\Line[ddd]\Linedots[rr]&&
\Linedots[dddddd]&1\\
 &4\Line[d]&4&4&&&&
 &4\Line[d]&4&4&&&\\
 &4\Line[d]&4&4&&&&
 &4\Line[d]&{\framebox{4}}&4&&&\\
&{\framebox{2}} \Line[r]\Line[d]&{\ast}\Line[rr]\Line[d]&&&&&
&5\Line[r]\Line[d]&{\ast}\Line[rr]\Line[d]&&&&\\
&3\Line[r]\Line[d]&3\Line[r]&\Line[dd]&&&&
&7\Line[d]\Line[r]&7\Line[r]&\Line[dd]&&&\\
ny&3\Line[d]&3&&&&&
  &7\Line[d]&7&&&&&ny\\
ny{+}1&\Line[rr]&&\Linedots[rrr]&&&&
&\Line[rr]&&\Linedots[rrr]&&&&ny{+}1\\
}$$}
\caption{Example of values of {\tt RightEdge} (left diagram) and
{\tt LowerEdge} (right diagram), respectively. Entry $\ast$ is indexed by
{\tt [2,4]}. It is not covered already since neither
$2<{\tt RightEdge[1,4]}=2$ nor $4<{\tt LowerEdge[2,3]}=4$.\label{fig:square}}
\end{figure}
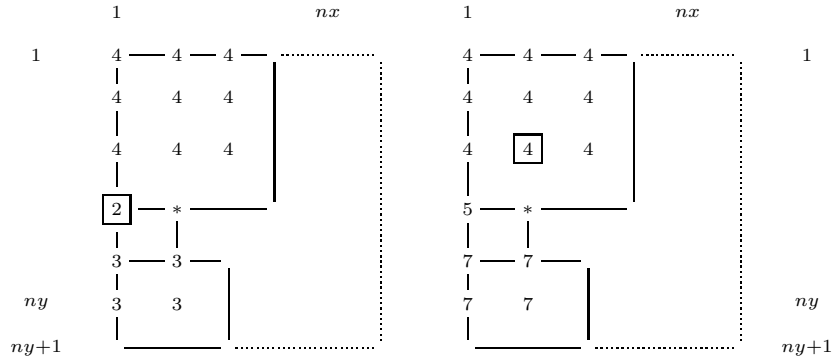

To illustrate the \almazero{} programming style and the
use of formulas as programs approach for program verification,
we consider now the following variant of a problem from 
\citeasnoun[pages 46-60]{Hon70}.
\\\\
{\it Squares in the rectangle.\/} 
Partition an integer sized $nx\times ny$ rectangle into given squares
$S_1,\dots,S_m$ of integer sizes $s_1,\dots,s_m$.
\\\\
We develop a solution that, in contrast to the one given in
\citeasnoun{ABPS98a}, is purely declarative.
To solve this problem we use a backtracking algorithm that fills in all the
cells of the rectangle one by one, starting with the left upper cell
and proceeding downward in the leftmost column, then the next column,
and so on. The algorithm checks for each cell whether it is already
covered by some square used to cover a previous cell. 
Given the order in which the cells are visited,
it suffices to inspect the left neighbour cell
and the upper neighbour cell 
(if these neighbours exist). This is done by the test
\begin{eqnarray*}
  \label{eq:test}
&&\lefteqn{{\tt ((1 < i)~AND~(i < RightEdge[i-1,j]))~OR~}}\\
&&{\tt ((1 < j)~AND~(j < LowerEdge[i, j-1]))}.
\end{eqnarray*}
Here \verb|[i,j]| is the index of the cell in question,
and \verb|RightEdge[i-1,j]| is the right edge of the square covering
the left neighbour (\verb|[i-1,j]|, provided \verb|i > 1|),
and  \verb|LowerEdge[i, j-1]| is the lower edge of the square covering
the upper neighbour (\verb|[i,j-1]|, provided \verb|j > 1|).
The cell under consideration is already covered if and only if 
the test succeeds.
If it is not covered, then the algorithm 
looks for a square not yet used, which is placed with its top-left corner
at \verb|[i,j]| provided the square fits within the rectangle. 
The algorithm backtracks when none of the available
squares can cover the cell under consideration
without sticking out of the rectangle. See Figure~\ref{fig:square}.

In test (\ref{eq:test}) we used the {\tt AND} and {\tt OR} connectives
instead of the ``;'' and {\tt ORELSE} constructs for the
following reason. In case all variables occurring in a test are instantiated,
some optimizations are in order. For example, it is not necessary
to backtrack within the test, disjunctions do not have to create
choice points, and so on. The use of {\tt AND} and {\tt OR}
enables the compiler to apply these optimizations.

Backtracking is implemented by a {\tt SOME} statement that checks for each
square whether it can be put to cover a given cell. The solution is returned
via two arrays {\tt posX} and {\tt posY} such that for square $S_k$ (of size
{\tt Sizes[k]}) {\tt posX[k]}, {\tt posY[k]} are the coordinates of its
top-left corner.

The two equations {\tt posX[k] = i} and {\tt posY[k] = j} are used both to
construct the solution and to prevent using an already placed square again at a
different place.

The declaration of the variables {\tt posX} and {\tt posY} as {\tt MIX} 
parameters allows us to use the program both to check a given 
solution or to complete a partial solution.

{\small 
\begin{verbatim}
TYPE SquaresVector = ARRAY [1..M] OF INTEGER;

PROCEDURE Squares(Sizes:SquaresVector, MIX posX, posY:SquaresVector);

VAR RightEdge,LowerEdge: ARRAY [1..NX],[1..NY] OF INTEGER;
    i,i1, j,j1, k: INTEGER;

BEGIN
  FOR i := 1 TO NX DO
    FOR j := 1 TO NY DO
      IF NOT                         (* cell [i,j] already covered? *)
         (((1 < i) AND (i < RightEdge[i-1,j])) OR
         ((1 < j) AND (j < LowerEdge[i, j-1])))    
      THEN 
        SOME k := 1 TO M DO
          PosX[k] = i;
          PosY[k] = j;                    (* square k already used? *)
          Sizes[k] + i <= NX + 1;
          Sizes[k] + j <= NY + 1;                 (* square k fits? *)
          FOR i1 := 1 TO Sizes[k] DO
            FOR j1 := 1 TO Sizes[k] DO
              RightEdge[i+i1-1,j+j1-1] = i+Sizes[k];
              LowerEdge[i+i1-1,j+j1-1] = j+Sizes[k] 
            END                          (* complete administration *)
          END
        END
      END
    END
  END
END Squares;

\end{verbatim}
}

\def\LE{{\tt {LowerEdge}}}
\def\RE{{\tt {RightEdge}}}
\def\pX{{\tt {PosX}}}
\def\pY{{\tt {PosY}}}
\def\Si{{\tt {Sizes}}}

This program is declarative and consequently has a dual
reading as the formula

\begin{eqnarray*}
&&\fa i\in[1..nx]~\fa j\in[1..ny]\\
&&\neg(1<i<\RE[i{-}1,j] \vee 1<j<\LE[i,j{-}1]) \ra \\
&&\te k\in[1..m]~\phi(i,j,k),
\end{eqnarray*}

\NI
where $\phi(i,j,k)$ is the formula

\begin{eqnarray*}
&&{\pX[k]=i}\wedge{\pY[k]=j}\wedge\\
&&{\Si[k]{+}i\leq nx{+}1}\wedge{\Si[k]{+}j\leq ny{+}1}\wedge\psi(i,j,k)
\end{eqnarray*}

\NI
and $\psi(i,j,k)$ is the formula

\begin{eqnarray*}
&&\fa i'\in[1..\Si(k)]~\fa j'\in[1..\Si(k)]\\
&&{\RE[i{+}i'{-}1,j{+}j'{-}1]=i{+}\Si[k]}\wedge\\
&&{\LE[i{+}i'{-}1,j{+}j'{-}1]=j{+}\Si[k]}
\end{eqnarray*}

\NI
This dual reading of the program entails over the standard
interpretation the formula

\begin{eqnarray}
&&\fa i\in[1..nx]~\fa j\in[1..ny]~\te k\in[1..m]\nonumber\\
&&{\pX[k]\leq i<\pX[k]{+}\Si[k]\leq nx{+}1} \wedge\nonumber\\
&&{\pY[k]\leq j<\pY[k]{+}\Si[k]\leq ny{+}1}\label{prob:spec}
\end{eqnarray}
\NI
expressing that every cell is covered by a square. 
The entailment is not trivial, but can be made completely rigorous.
The proof uses arithmetic,
in particular induction on lexicographically ordered pairs $(i,j)$.
This entailment actually means that the program satisfies 
its specification, that is, 
if the computation is successful, then a partition is found (and can
be read off from $\pX[k]$ and $\pY[k]$). The latter fact relies
on the Soundness Theorem~\ref{thm:soundn}. 

Conversely, assuming that the surfaces of the squares sum up exactly to
the surface of the rectangle, the specification (\ref{prob:spec}) entails 
the formula corresponding to the program, with suitable values for $\RE,\LE$.
Furthermore, the absence of
errors can be established by lexicographic induction. This ensures that
the computation tree is always determined. 
By the Completeness Theorem~\ref{thm:completeness}, one always gets an answer.
If this answer is negative, that is, if the computation tree is failed,
then by the Soundness Theorem~\ref{thm:soundn} the formula corresponding
to the program cannot be satisfied, and hence (\ref{prob:spec}) 
cannot be satisfied.

\section{Current and Future Work}

The work here presented can be pursued in a number of directions.
We listed here the ones that seem to us most natural.

\paragraph{Recursive procedures}
The extension of the treatment of non-recursive procedures in
Subsection \ref{subsec: nonrec} to the case of recursive procedures is
far from obvious.  It requires an extension of the computation
mechanism to one with possible non-terminating behaviour. This could
be done along the lines of \citeasnoun{AD94} where the
SLDNF-resolution of logic programs with negation is presented in a top
down, non-circular way.

Also, on the semantic level several choices arise, much like in the
case of logic programming, and the corresponding soundness and
completeness results that provide a match between the computation
mechanism and semantics need to be reconsidered from scratch.

\paragraph{Constraints}
As already said in Subsection \ref{subsec:clp}, the formulas as
programs approach can be seen as a special case of constraint logic
programming, though with a full first-order syntax. It is natural to
extend our approach by allowing constraints, so arbitrary atoms that
have no definition in the sense of Subsection \ref{subsec: nonrec}.
The addition of constraints will require on the computation mechanism
level use of a constraint store and special built-in procedures that
approximate the satisfiability test for conjunctions of constraints.

\paragraph{Automated Verification}
The correctness proof presented in Section \ref{sec:squares} was
carried out manually. It boils down to a proof of validity of an
implication between two formulas, This proof is based on an
lexicographic ordering os it should be possible to mechanize this
proof. This would lead a fully mechanized correctness proof of the
\almazero{} program considered there.

\paragraph{Relation to Dynamic Predicate Logic}
In \citeasnoun{GS91} an alternative ``input-output'' semantics of
first-order logic is provided.  In this semantics both the connectives
and the quantifiers obtain a different, dynamic, interpretation that
better suits their use for natural language analysis.
This semantic is highly nondeterministic due to its treatment of
existential quantifiers and it does not take into account a possibility
of errors.

It is natural to investigate the precise connection between this
semantics and our formulas as programs approach. A colleague of us,
Jan van Eijck, has recently undertook this study. Also, it would be
useful to clarify to what extent our approach can be of use for
linguistic analysis, both as a computation mechanism and as a means
for capturing errors in discourse analysis.

\paragraph{Absence of abnormal termination}
Another natural line of research deals with the improvements of the
computation mechanism in the sense of limiting the occurrence of 
errors while retaining soundness. In  Appendix, Subsections~\ref{ssec:libneg}
and \ref{ssec:libimp} we consider two such possibilities but several
other options arise. Also, it is useful to provide 
sufficient syntactic criteria that for a formula
guarantee absence of abnormal termination. This work is naturally related
to a research on verification of \almazero{} programs.

\renewcommand{\thebibliography}[1]{
  \section*{References}%
  \list{%
      \arabic{enumi}.\hfill
    }{%
      \topsep0pt\parskip0pt\partopsep0pt
      \settowidth\labelwidth{#1.}%
      \leftmargin\labelwidth
      \advance\leftmargin\labelsep
      \usecounter{enumi}%
      \itemsep.3\baselineskip
      \parsep0pt
    }
  \def\newblock{\hskip .11em plus .33em minus .07em}%
  \sloppy\clubpenalty4000\widowpenalty4000
  \sfcode`\.=1000\relax}

\section*{Acknowledgements}
We would like to thank Jan van Eijck and David Scott Warren for a
number of helpful suggestions.

\bibliographystyle{../art/agsm}
\bibliography{/ufs/apt/book-ao-2nd/apt,/ufs/apt/esprit/esprit,/ufs/apt/esprit/chapter3,/ufs/apt/bib/clp2,/ufs/apt/bib/clp1,/ufs/apt/book-lp/man1,/ufs/apt/book-lp/man2,/ufs/apt/book-lp/man3,/ufs/apt/book-lp/ref1,/ufs/apt/book-lp/ref2,/ufs/apt/bib/a0}

\section{Appendix}
 
\subsection{Proof of the Soundness Theorem~\ref{thm:soundn}}
\label{ssec:soundn}
The proof proceeds by induction on the lexicographic ordering on 
formulas which is defined in Definition~\ref{def:tree}. 
We carefully go through all inductive cases.

\begin{enumerate}
\item The case of the empty conjunction is trivial.
\item The first three of the four cases concerning atom $A$ are obvious.
It remains to deal with the last case, where atom $A$ 
is an $\alpha$-assignment $s = t$. Then either
$s$ or $t$ is a variable which is not $\alpha$-closed,
say $s \equiv x$ with $x$ not $\alpha$-closed and $t$
$\alpha$-closed. The symmetrical case is analogous. The tree
$\B{{x=t}\wedge\psi}_\alpha$ is, apart from the root
of degree one, identical to $\B{\psi}_{\alpha\cup\{x/t^\alpha\}}$.

If $\B{\psi}_{\alpha\cup\{x/t^\alpha\}}$ contains a success
leaf labelled by $\beta$, then by the induction hypothesis
$\fa(\psi^{\beta})$ is true. Since $t$ is $\alpha$-closed
and $\beta$ extends $\alpha\cup\{x/t^\alpha\}$,
we have $(x=t)^{\beta} \equiv (x^{\beta} = t^{\beta})
\equiv (t^{\alpha} = t^{\alpha})$.
The last formula is true, so also
$\fa (({x=t}\wedge\psi)^{\beta})$ is true.
 
If $\B{\psi}_{\alpha\cup\{x/t^\alpha\}}$ is failed, 
then by the induction hypothesis
$\te(\psi^{\alpha\cup\{x/t^\alpha\}})$ is false.
Note again that $t$ is $\alpha$-closed and let
$x,x_1,\ldots,x_n$ be all the free variables of $\psi$ that
are not in the domain of $\alpha$. (If $n=0$ or if $x$ does
not occur in $\psi$, then the argument is even simpler.)
Then we have $\te(({x=t} \wedge\psi)^{\alpha}) \equiv
\exists x,x_1,\ldots,x_n \;(x=t^{\alpha}\wedge
\psi^\alpha(x,x_1,\ldots,x_n))$, which is logically equivalent to
$\exists x_1,\ldots,x_n \; \psi^\alpha(t^\alpha,x_1,\ldots,x_n)) 
\equiv \te(\psi^{\alpha\cup\{x/t^\alpha\}})$.
It follows that $\te(({x=t} \wedge\psi)^{\alpha})$ is also false.
\item The case of $(\phi_1\vee\phi_2)\wedge\psi$ uses
the distributive law and the induction hypothesis applied
to the the lexicographically 
smaller formulas $\phi_1\wedge\psi$ and $\phi_2\wedge\psi$.
\item The case of $(\phi_1\wedge\phi_2)\wedge\psi$  uses
the associativity of conjunction and the induction hypothesis applied
to the the lexicographically smaller formulas $\phi_1$ and $\phi_2\wedge\psi$.
\item The case of $(\phi_1\ra\phi_2)\wedge\psi$  uses
the logical equivalence of
$\phi_1 \ra\phi_2$ and $\neg\phi_1 \vee \phi_2$.
If formula $\phi_1$ is $\alpha$-closed and $\B{\phi}_\alpha$
is failed, then the argument is similar to the corresponding
case of $\neg\phi \wedge \psi$ in the next case.
The other case can be dealt with by applying the 
induction hypothesis to $\phi_1 \A (\phi_2\wedge\psi)$.
\item For $\neg\phi \wedge\psi$ 
we distinguish three cases with respect to $\phi$.
\begin{itemize}
\item Formula $\phi$ is $\alpha$-closed and $\B{\phi}_\alpha$
is failed. Then, by the induction hypothesis,
$\te(\phi^\alpha)$ is false, so $\fa(\neg\phi^\alpha)$ is true.
Since $\B{\neg\phi \wedge\psi}_\alpha$ is, apart from the root
of degree one, identical to $\B{\psi}_\alpha$, we apply
the induction hypothesis to $\psi$. If $\B{\psi}_\alpha$ is failed, 
then $\te(\psi^\alpha)$ is false, and hence
$\te((\neg\phi \wedge\psi)^\alpha)$ is false.
If $\B{\psi}_\alpha$ contains a success leaf $\beta$
then $\beta$ extends $\alpha$ and $\fa(\psi^\beta)$ is true.
Note that $\fa(\neg\phi^\alpha)$ implies $\fa(\neg\phi^\gamma)$,
for any $\gamma$ extending $\alpha$, \emph{even if $\phi$ is
not $\alpha$-closed}.
It follows that $\fa((\neg\phi \wedge\psi)^\beta)$ is true.
Observe that we did not use the fact that $\phi$ is $\alpha$-closed.
So the proof remains valid under the first relaxation described in 
Subsection~\ref{ssec:libneg}.
\item Formula $\phi$ is $\alpha$-closed and $\B{\phi}_\alpha$
contains at least one success leaf, labelled by an extension $\beta$
of $\alpha$. The tree $\B{\neg\phi \wedge\psi}_\alpha$ consists of a root
and a failure leaf in this case,
so we have to show that $\te((\neg\phi \wedge\psi)^\alpha)$ is false.
By the induction hypothesis, $\fa(\phi^{\beta})$ is true,
and hence $\fa(\phi^\alpha)$ is true, as 
$\beta$ is an extension of $\alpha$ and $\phi$ is $\alpha$-closed.
\emph{This implication also holds if $\phi$ is
not $\alpha$-closed, provided that $\beta$ does not contain
any pair $x/d$ where $x$ is free in $\phi^\alpha$.}
Consequently, $\te(\neg\phi^\alpha)$ is false and hence also
$\te((\neg\phi \wedge\psi)^\alpha)$ is false.
Observe that the proof remains valid under the second relaxation 
described in Subsection~\ref{ssec:libneg}.
\item In all other cases there is nothing to prove as 
$\B{\neg\phi \wedge\psi}_\alpha$ has then only $\err$ leaves.
\end{itemize}
\item For the case $\exists x\;\phi \wedge\psi$,
assume that $x$ is fresh with respect to $\psi$ and some valuation
$\alpha$. It is convenient to make the possible occurrence of $x$
in $\phi$ explicit by writing $\phi(x)$ for $\phi$.
Recall that apart form the root of degree one
$\B{\exists x\;\phi(x) \wedge\psi}_\alpha$ is identical to
$\B{\phi(x)\wedge\psi}_\alpha$.

Assume $\B{\phi(x)\wedge\psi}_\alpha$ contains a success leaf
labelled by $\beta$. By applying the induction hypothesis
to the lexicographically smaller formula $\phi(x)\wedge\psi$ we get that 
$\fa((\phi(x)\wedge\psi)^{\beta})$ is true. 
It follows that 
$\fa((\exists x\;\phi(x) \wedge\psi)^{\beta})$ is true.
Some minor technicalities have been left to the reader here: the
case in which $x$ does not occur in the domain of $\beta$
has to be settled by applying 
$(\forall x\;\phi(x))\ra (\exists x\;\phi(x))$
and not by inferring $\exists x\;\phi(x)$ from $\phi(x^{\beta})$.

Assume $\B{\phi(x)\wedge\psi}_\alpha$ is failed.
Then, again by the induction hypothesis,
$\te((\phi(x)\wedge\psi)^{\alpha})$ is false.
Since $x$ does occur neither in $\alpha$, nor in $\psi$, it follows
that $\te((\exists x\;\phi(x) \wedge\psi)^{\alpha})$ is false.
\HB
\end{enumerate}
\NI
\subsection{Proof of the Restricted Completeness Theorem~\ref{thm:completeness}}
\label{ssec:completeness}

\NI (i) Suppose by contradiction that $\B{\phi}_\alpha$ is not
successful. Since this tree is determined, it is failed.  By the
Soundness Theorem~\ref{thm:soundn} $\te (\phi^{\alpha})$ is false
which is a contradiction. 
\\[2mm] 
\NI 
(ii) Suppose by contradiction
that $\B{\phi}_\alpha$ is not failed. Since this tree is determined,
it is successful.  By the Soundness Theorem~\ref{thm:soundn} for some
$\beta$ that extends $\alpha$ we have that $\fa (\phi^{\beta})$ is
true.  This is a contradiction since the falsity of $\te
(\phi^{\alpha})$ is equivalent to the truth $\fa (\neg \phi^{\alpha})$
that implies the truth of $\fa (\neg \phi^{\beta})$.  \HB

\subsection{More liberal negation}\label{ssec:libneg}
In this subsection we show how in Definition~\ref{def:tree}
the restriction ``$\phi$ is $\alpha$-closed'' in the case of the tree 
$\B{\neg\phi \wedge\psi}_\alpha$ can be relaxed without losing soundness.
There are basically two such relaxations.

First, observe by means of example that $\neg({0=1}\wedge{x=y})$ is true, 
independent of the values of $x$ and $y$.
This observation can be generalized as follows.
If $\B{\phi}_\alpha$ is failed,
then $\B{\neg\phi \wedge\psi}_\alpha$ can be defined as the tree
with a root of degree one and $\B{\psi}_\alpha$ as its subtree,
\emph{even if $\phi$ is not $\alpha$-closed.}
In the proof of the Soundness Theorem we already accommodated
for this relaxation, see Subsection~\ref{ssec:soundn}. 

Second, observe that the dual phenomenon also exists:
$\neg({0=0}\vee{x=y})$ is false, independent of the values of $x$ and $y$.
More generally, if $\B{\phi}_\alpha$ contains a success leaf $\beta$
not containing any pair $x/d$ with $x$ free in $\phi^\alpha$,
then $\B{\neg\phi \wedge\psi}_\alpha$ can be defined as the tree
with a root of degree one and a failure leaf as its son,
\emph{even if $\phi$ is not $\alpha$-closed.}

\subsection{More liberal implication}\label{ssec:libimp}

The first and the second relaxation above can be both applied to the 
computation tree $\B{(\phi_1\ra\phi_2)\wedge\psi}_\alpha$,
the first to the case in which the tree $\B{\phi_1}_\alpha$ is failed,
and the second to the case in which the tree $\B{\phi_1}_\alpha$
contains a success leaf 
not containing any pair $x/d$ with $x$ free in $\phi_1^\alpha$.

There are several other ways to liberalize implication.
The aim is to be more complete, that is, to yield more determined
computation trees (without losing soundness, of course).

As a first example, consider the following computation tree:

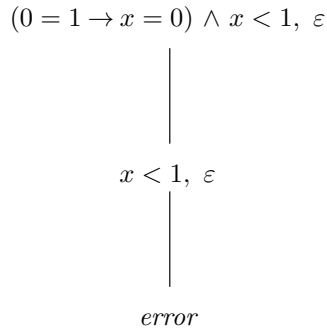
\begin{center}
\begin{picture}(0,0)%
\includegraphics{aptbezem19.pstex}%
\end{picture}%
\setlength{\unitlength}{3947sp}%
\begingroup\makeatletter\ifx\SetFigFont\undefined
\def\x#1#2#3#4#5#6#7\relax{\def\x{#1#2#3#4#5#6}}%
\expandafter\x\fmtname xxxxxx\relax \def\y{splain}%
\ifx\x\y   
\gdef\SetFigFont#1#2#3{%
  \ifnum #1<17\tiny\else \ifnum #1<20\small\else
  \ifnum #1<24\normalsize\else \ifnum #1<29\large\else
  \ifnum #1<34\Large\else \ifnum #1<41\LARGE\else
     \huge\fi\fi\fi\fi\fi\fi
  \csname #3\endcsname}%
\else
\gdef\SetFigFont#1#2#3{\begingroup
  \count@#1\relax \ifnum 25<\count@\count@25\fi
  \def\x{\endgroup\@setsize\SetFigFont{#2pt}}%
  \expandafter\x
    \csname \romannumeral\the\count@ pt\expandafter\endcsname
    \csname @\romannumeral\the\count@ pt\endcsname
  \csname #3\endcsname}%
\fi
\fi\endgroup
\begin{picture}(2539,2034)(2631,-2113)
\put(3901,-2086){\makebox(0,0)[b]{\smash{\SetFigFont{10}{13.2}{rm}{\em error}}}}
\put(3901,-1186){\makebox(0,0)[b]{\smash{\SetFigFont{10}{13.2}{rm}$x < 1, \ \varepsilon$}}}
\put(3901,-211){\makebox(0,0)[b]{\smash{\SetFigFont{10}{13.2}{rm}$(0 = 1 \ra x = 0) \A x < 1, \ \varepsilon$}}}
\end{picture}

\end{center}

This computation tree is not determined. In contrast,
using the equivalence of
${\phi_1 \ra \phi_2}$ and ${\neg\phi_1 \vee \phi_2}$,
we get the following computation tree
which is determined,
as it contains a success leaf $\{x/0\}$:
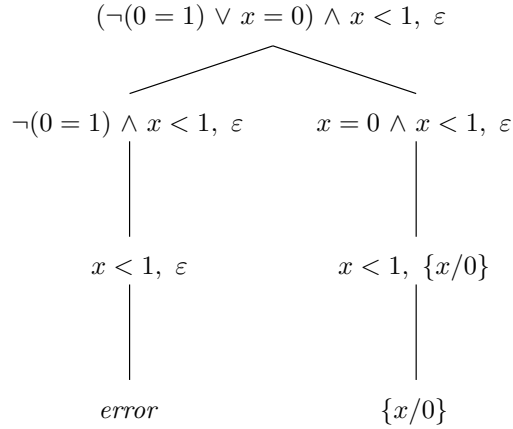
\begin{center}
\begin{picture}(0,0)%
\includegraphics{aptbezem20.pstex}%
\end{picture}%
\setlength{\unitlength}{3947sp}%
\begingroup\makeatletter\ifx\SetFigFont\undefined
\def\x#1#2#3#4#5#6#7\relax{\def\x{#1#2#3#4#5#6}}%
\expandafter\x\fmtname xxxxxx\relax \def\y{splain}%
\ifx\x\y   
\gdef\SetFigFont#1#2#3{%
  \ifnum #1<17\tiny\else \ifnum #1<20\small\else
  \ifnum #1<24\normalsize\else \ifnum #1<29\large\else
  \ifnum #1<34\Large\else \ifnum #1<41\LARGE\else
     \huge\fi\fi\fi\fi\fi\fi
  \csname #3\endcsname}%
\else
\gdef\SetFigFont#1#2#3{\begingroup
  \count@#1\relax \ifnum 25<\count@\count@25\fi
  \def\x{\endgroup\@setsize\SetFigFont{#2pt}}%
  \expandafter\x
    \csname \romannumeral\the\count@ pt\expandafter\endcsname
    \csname @\romannumeral\the\count@ pt\endcsname
  \csname #3\endcsname}%
\fi
\fi\endgroup
\begin{picture}(3892,2637)(3047,-2716)
\put(4201,-886){\makebox(0,0)[b]{\smash{\SetFigFont{10}{13.2}{rm}$ \neg (0 = 1) \A x < 1, \ \varepsilon$}}}
\put(4276,-1786){\makebox(0,0)[b]{\smash{\SetFigFont{10}{13.2}{rm}$x < 1, \ \varepsilon$}}}
\put(4201,-2686){\makebox(0,0)[b]{\smash{\SetFigFont{10}{13.2}{rm}{\em error}}}}
\put(6001,-886){\makebox(0,0)[b]{\smash{\SetFigFont{10}{13.2}{rm}$x = 0 \A x < 1, \ \varepsilon$}}}
\put(6001,-1786){\makebox(0,0)[b]{\smash{\SetFigFont{10}{13.2}{rm}$x < 1, \ \C{x/0}$}}}
\put(6001,-2686){\makebox(0,0)[b]{\smash{\SetFigFont{10}{13.2}{rm}$\C{x/0}$}}}
\put(5101,-211){\makebox(0,0)[b]{\smash{\SetFigFont{10}{13.2}{rm}$(\neg (0 =1) \Or x = 0) \A  x < 1 , \ \varepsilon$}}}
\end{picture}

\end{center}

The above example shows that $\neg\phi_1 \vee \phi_2$ can be
``more complete'' than $\phi_1 \ra \phi_2$, although in some cases
the disjunction involves unnecessary branching in the computation tree. 
As an example of the latter phenomenon, compare the computation trees
for $({0=1}\ra{0=0})\wedge{\psi}$ and $({\neg (0=1)}\vee{0=0})\wedge{\psi}$:

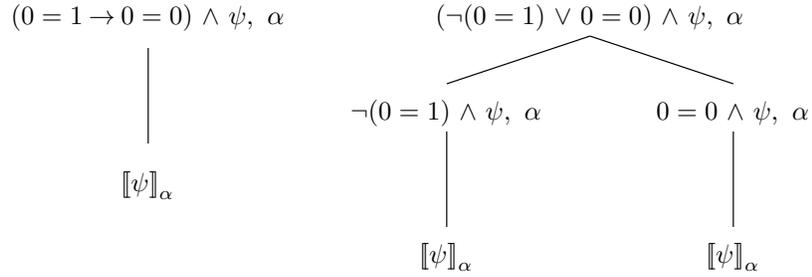
\begin{center}
\begin{picture}(0,0)%
\includegraphics{aptbezem21.pstex}%
\end{picture}%
\setlength{\unitlength}{3947sp}%
\begingroup\makeatletter\ifx\SetFigFont\undefined
\def\x#1#2#3#4#5#6#7\relax{\def\x{#1#2#3#4#5#6}}%
\expandafter\x\fmtname xxxxxx\relax \def\y{splain}%
\ifx\x\y   
\gdef\SetFigFont#1#2#3{%
  \ifnum #1<17\tiny\else \ifnum #1<20\small\else
  \ifnum #1<24\normalsize\else \ifnum #1<29\large\else
  \ifnum #1<34\Large\else \ifnum #1<41\LARGE\else
     \huge\fi\fi\fi\fi\fi\fi
  \csname #3\endcsname}%
\else
\gdef\SetFigFont#1#2#3{\begingroup
  \count@#1\relax \ifnum 25<\count@\count@25\fi
  \def\x{\endgroup\@setsize\SetFigFont{#2pt}}%
  \expandafter\x
    \csname \romannumeral\the\count@ pt\expandafter\endcsname
    \csname @\romannumeral\the\count@ pt\endcsname
  \csname #3\endcsname}%
\fi
\fi\endgroup
\begin{picture}(5467,1665)(2839,-1744)
\put(7576,-1711){\makebox(0,0)[b]{\smash{\SetFigFont{10}{13.2}{rm}$\B{\psi}_\alpha$}}}
\put(7576,-811){\makebox(0,0)[b]{\smash{\SetFigFont{10}{13.2}{rm}$0 = 0 \A \psi, \ \alpha$}}}
\put(3901,-211){\makebox(0,0)[b]{\smash{\SetFigFont{10}{13.2}{rm}$(0 = 1 \ra 0 = 0) \A \psi, \ \alpha$}}}
\put(3901,-1261){\makebox(0,0)[b]{\smash{\SetFigFont{10}{13.2}{rm}$\B{\psi}_\alpha$}}}
\put(5776,-1711){\makebox(0,0)[b]{\smash{\SetFigFont{10}{13.2}{rm}$\B{\psi}_\alpha$}}}
\put(6676,-211){\makebox(0,0)[b]{\smash{\SetFigFont{10}{13.2}{rm}$(\neg (0 = 1) \Or 0 = 0) \A  \psi, \ \alpha$}}}
\put(5776,-811){\makebox(0,0)[b]{\smash{\SetFigFont{10}{13.2}{rm}$ \neg (0 = 1) \A \psi,  \ \alpha$}}}
\end{picture}

\end{center}

As a second example, consider the computation tree
$\B{({x=0}\ra{x<1}}_\empval$, which is not determined since $x=0$ is
not $\empval$-closed:

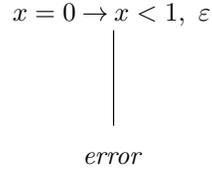
\begin{center}
\begin{picture}(0,0)%
\includegraphics{aptbezem22.pstex}%
\end{picture}%
\setlength{\unitlength}{3947sp}%
\begingroup\makeatletter\ifx\SetFigFont\undefined
\def\x#1#2#3#4#5#6#7\relax{\def\x{#1#2#3#4#5#6}}%
\expandafter\x\fmtname xxxxxx\relax \def\y{splain}%
\ifx\x\y   
\gdef\SetFigFont#1#2#3{%
  \ifnum #1<17\tiny\else \ifnum #1<20\small\else
  \ifnum #1<24\normalsize\else \ifnum #1<29\large\else
  \ifnum #1<34\Large\else \ifnum #1<41\LARGE\else
     \huge\fi\fi\fi\fi\fi\fi
  \csname #3\endcsname}%
\else
\gdef\SetFigFont#1#2#3{\begingroup
  \count@#1\relax \ifnum 25<\count@\count@25\fi
  \def\x{\endgroup\@setsize\SetFigFont{#2pt}}%
  \expandafter\x
    \csname \romannumeral\the\count@ pt\expandafter\endcsname
    \csname @\romannumeral\the\count@ pt\endcsname
  \csname #3\endcsname}%
\fi
\fi\endgroup
\begin{picture}(1877,1059)(2962,-2713)
\put(3901,-1786){\makebox(0,0)[b]{\smash{\SetFigFont{10}{13.2}{rm}$x = 0 \ra x < 1, \ \varepsilon$}}}
\put(3901,-2686){\makebox(0,0)[b]{\smash{\SetFigFont{10}{13.2}{rm}{\em error}}}}
\end{picture}

\end{center}

One would like to have this tree succeed with $\{x/0\}$. (The fact that
$\{x/1\}$ is also a solution is beyond the scope of our method,
since $\B{\neg (x=0)}_\empval$ is not determined.)
For this the equivalence of ${\phi_1 \ra \phi_2}$ and
${\neg\phi_1 \vee \phi_2}$ does not help, as the computation tree
$\B{\neg (x=0)\vee{x<1}}_\empval$ is not determined either:

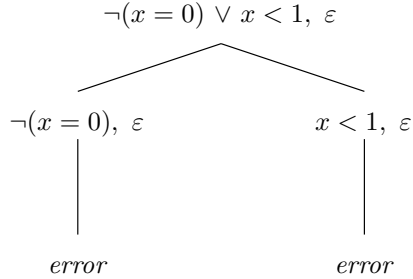
\begin{center}
\begin{picture}(0,0)%
\includegraphics{aptbezem23.pstex}%
\end{picture}%
\setlength{\unitlength}{3947sp}%
\begingroup\makeatletter\ifx\SetFigFont\undefined
\def\x#1#2#3#4#5#6#7\relax{\def\x{#1#2#3#4#5#6}}%
\expandafter\x\fmtname xxxxxx\relax \def\y{splain}%
\ifx\x\y   
\gdef\SetFigFont#1#2#3{%
  \ifnum #1<17\tiny\else \ifnum #1<20\small\else
  \ifnum #1<24\normalsize\else \ifnum #1<29\large\else
  \ifnum #1<34\Large\else \ifnum #1<41\LARGE\else
     \huge\fi\fi\fi\fi\fi\fi
  \csname #3\endcsname}%
\else
\gdef\SetFigFont#1#2#3{\begingroup
  \count@#1\relax \ifnum 25<\count@\count@25\fi
  \def\x{\endgroup\@setsize\SetFigFont{#2pt}}%
  \expandafter\x
    \csname \romannumeral\the\count@ pt\expandafter\endcsname
    \csname @\romannumeral\the\count@ pt\endcsname
  \csname #3\endcsname}%
\fi
\fi\endgroup
\begin{picture}(3354,1734)(3308,-1813)
\put(4201,-886){\makebox(0,0)[b]{\smash{\SetFigFont{10}{13.2}{rm}$ \neg (x = 0), \ \varepsilon$}}}
\put(6001,-886){\makebox(0,0)[b]{\smash{\SetFigFont{10}{13.2}{rm}$x < 1, \ \varepsilon$}}}
\put(4201,-1786){\makebox(0,0)[b]{\smash{\SetFigFont{10}{13.2}{rm}{\em error}}}}
\put(6001,-1786){\makebox(0,0)[b]{\smash{\SetFigFont{10}{13.2}{rm}{\em error}}}}
\put(5101,-211){\makebox(0,0)[b]{\smash{\SetFigFont{10}{13.2}{rm}$\neg (x = 0) \Or x < 1, \ \varepsilon$}}}
\end{picture}

\end{center}

Note that the left subtree ends with $\err$ since $\B{x=0}$
succeeds with $\{x/0\}$. Liberal negation does not help us
any further here.

In order to have $\B{({x=0}\ra{x<1}}_\empval$ succeed it is necessary
to transfer the valuation of the success leaf of the antecedent,
i.e.\ $\{x/0\}$, to the consequent. Thus we are tempted to consider
$\neg\phi_1 \vee (\phi_1 \wedge \phi_2)$ as a more useful logical
equivalent of ${\phi_1 \ra \phi_2}$ than ${\neg\phi_1 \vee \phi_2}$.
The conjunction $\phi_1 \wedge \phi_2$ has the desired effect
on the transfer of valuations. Indeed the following computation tree
is successful:

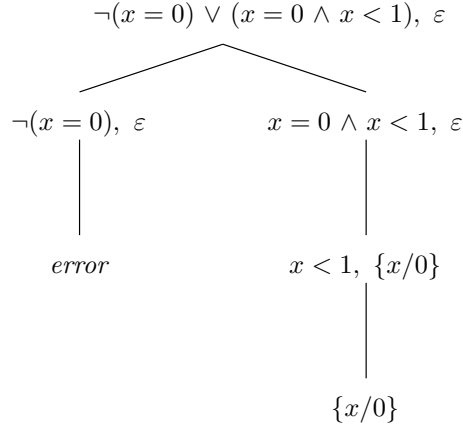
\begin{center}
\begin{picture}(0,0)%
\includegraphics{aptbezem24.pstex}%
\end{picture}%
\setlength{\unitlength}{3947sp}%
\begingroup\makeatletter\ifx\SetFigFont\undefined
\def\x#1#2#3#4#5#6#7\relax{\def\x{#1#2#3#4#5#6}}%
\expandafter\x\fmtname xxxxxx\relax \def\y{splain}%
\ifx\x\y   
\gdef\SetFigFont#1#2#3{%
  \ifnum #1<17\tiny\else \ifnum #1<20\small\else
  \ifnum #1<24\normalsize\else \ifnum #1<29\large\else
  \ifnum #1<34\Large\else \ifnum #1<41\LARGE\else
     \huge\fi\fi\fi\fi\fi\fi
  \csname #3\endcsname}%
\else
\gdef\SetFigFont#1#2#3{\begingroup
  \count@#1\relax \ifnum 25<\count@\count@25\fi
  \def\x{\endgroup\@setsize\SetFigFont{#2pt}}%
  \expandafter\x
    \csname \romannumeral\the\count@ pt\expandafter\endcsname
    \csname @\romannumeral\the\count@ pt\endcsname
  \csname #3\endcsname}%
\fi
\fi\endgroup
\begin{picture}(3646,2637)(3008,-2716)
\put(3901,-886){\makebox(0,0)[b]{\smash{\SetFigFont{10}{13.2}{rm}$ \neg (x = 0), \ \varepsilon$}}}
\put(3901,-1786){\makebox(0,0)[b]{\smash{\SetFigFont{10}{13.2}{rm}{\em error}}}}
\put(5701,-886){\makebox(0,0)[b]{\smash{\SetFigFont{10}{13.2}{rm}$x = 0  \A x < 1, \ \varepsilon$}}}
\put(5701,-1786){\makebox(0,0)[b]{\smash{\SetFigFont{10}{13.2}{rm}$x < 1, \  \C{x/0}$}}}
\put(5701,-2686){\makebox(0,0)[b]{\smash{\SetFigFont{10}{13.2}{rm}$\C{x/0}$}}}
\put(5101,-211){\makebox(0,0)[b]{\smash{\SetFigFont{10}{13.2}{rm}$\neg (x = 0) \Or (x = 0  \A x < 1), \ \varepsilon$}}}
\end{picture}

\end{center}

The combination of $\neg\phi_1 \vee (\phi_1 \wedge \phi_2)$
for $\phi_1 \ra \phi_2$
with liberal negation yields the following computation tree for
$\B{(({x=0 \wedge x=1})\ra{0=1})\wedge{\psi}}_\empval$:

\begin{center}
\begin{picture}(0,0)%
\includegraphics{aptbezem25.pstex}%
\end{picture}%
\setlength{\unitlength}{3947sp}%
\begingroup\makeatletter\ifx\SetFigFont\undefined
\def\x#1#2#3#4#5#6#7\relax{\def\x{#1#2#3#4#5#6}}%
\expandafter\x\fmtname xxxxxx\relax \def\y{splain}%
\ifx\x\y   
\gdef\SetFigFont#1#2#3{%
  \ifnum #1<17\tiny\else \ifnum #1<20\small\else
  \ifnum #1<24\normalsize\else \ifnum #1<29\large\else
  \ifnum #1<34\Large\else \ifnum #1<41\LARGE\else
     \huge\fi\fi\fi\fi\fi\fi
  \csname #3\endcsname}%
\else
\gdef\SetFigFont#1#2#3{\begingroup
  \count@#1\relax \ifnum 25<\count@\count@25\fi
  \def\x{\endgroup\@setsize\SetFigFont{#2pt}}%
  \expandafter\x
    \csname \romannumeral\the\count@ pt\expandafter\endcsname
    \csname @\romannumeral\the\count@ pt\endcsname
  \csname #3\endcsname}%
\fi
\fi\endgroup
\begin{picture}(5331,2634)(2139,-2713)
\put(3601,-886){\makebox(0,0)[b]{\smash{\SetFigFont{10}{13.2}{rm}$\neg (x = 0 \A x = 1) \A \psi, \ \varepsilon$}}}
\put(6001,-886){\makebox(0,0)[b]{\smash{\SetFigFont{10}{13.2}{rm}$x = 0  \A x = 1 \A 0 = 1 \A \psi, \ \varepsilon$}}}
\put(6001,-1786){\makebox(0,0)[b]{\smash{\SetFigFont{10}{13.2}{rm}$x =1 \A 0 = 1 \A \psi,  \  \C{x/0}$}}}
\put(6001,-2686){\makebox(0,0)[b]{\smash{\SetFigFont{10}{13.2}{rm}{\em fail}}}}
\put(3601,-1786){\makebox(0,0)[b]{\smash{\SetFigFont{10}{13.2}{rm}$\B{\psi}_\varepsilon$}}}
\put(5101,-211){\makebox(0,0)[b]{\smash{\SetFigFont{10}{13.2}{rm}$(\neg (x = 0 \A x = 1) \Or (x = 0  \A x = 1 \A 0 = 1)) \A \psi, \ \varepsilon$}}}
\end{picture}

\end{center}

Note that the ``guard''
${x=0}\wedge{x=1}$ prevents $\B{\psi}_\empval$ to be computed twice,
even if we replace $0=1$ by $0=0$. 

On the other hand, this guard
also prevents successful computations, such as in
$\B{(({x=0 \wedge x=1})\ra{x=0})\wedge{x<1}}_\empval$,
where the solution $\{x/0\}$ is missed when 
$\neg\phi_1 \vee (\phi_1 \wedge \phi_2)$ is used instead of
${\neg\phi_1 \vee \phi_2}$ for ${\phi_1 \ra \phi_2}$.

The above example shows that $\neg\phi_1 \vee (\phi_1 \wedge \phi_2)$
is not always ``more complete'' than ${\neg\phi_1 \vee \phi_2}$.
Thus we are led to consider 
$\neg\phi_1 \vee \phi_2 \vee (\phi_1 \wedge \phi_2)$ as a third
logical equivalent of ${\phi_1 \ra \phi_2}$, in an attempt to
collect all the successes of ${\neg\phi_1 \vee \phi_2}$ and
$\neg\phi_1 \vee (\phi_1 \wedge \phi_2)$. Indeed this works
for the successes, but not for the failures, as the following delicate
example shows.

Consider $\B{\phi_1 \ra \phi_2}_\empval$ with ${0=0}\vee{x<1}$ for
$\phi_1$ and $0=1$ for $\phi_2$. Then both $\B{\neg\phi_1}_\empval$ and
$\B{\phi_2}_\empval$ are failed, but $\B{\phi_1 \wedge \phi_2}_\empval$
has an $\err$ leaf due to the disjunct $x<1$.
This means that $\B{\neg\phi_1\vee \phi_2}_\empval$ is failed,
whereas neither $\B{\neg\phi_1\vee (\phi_1\wedge\phi_2)}_\empval$
nor $\B{\neg\phi_1\vee \phi_2 \vee (\phi_1\wedge\phi_2)}_\empval$ 
is determined.

From the above we can draw the following conclusions:
\begin{itemize}
\item Liberal negation is always an improvement for implication;

\item For finding successes, use ${\phi_1 \ra \phi_2}\equiv
{\neg\phi_1\vee \phi_2 \vee (\phi_1\wedge\phi_2)}$;

\item For failures, use ${\phi_1 \ra \phi_2}\equiv{\neg\phi_1\vee \phi_2}$.
\end{itemize}

\end{document}